\documentclass[paper,prd,reprint,preprintnumbers,notitlepage,nofootinbib]{revtex4-1}

\usepackage{graphicx,color}
\usepackage{amssymb,amsmath}

\begin{document}

\preprint{UTHEP-727, UTCCS-P-118}

\title{
Quark mass dependence of on-shell and half off-shell scattering amplitudes
from Bethe-Salpeter wave function inside the interaction range
}

\date{
\today
%Ver.11 Mar 29th 2019 by YN
%Ver.10 Mar 26th 2019 by YN
%Ver.09 Jan 22th 2019 by YN and TY
%Ver.08 Jan 15th 2019 by YN and TY
%Ver.07 Dec 26th 2018 by YN and TY
%Ver.06 Dec 21th 2018 by YN and TY
%Ver.05 Dec 11th 2018 by YN
%Ver.04 Dec  3rd 2018 by TY
%Ver.03 Nov  1st 2018 by YN
%Ver.02 Oct 29th 2018 by YN
%Ver.01 Oct 23rd 2018 by YN
%Ver.00 Oct 15th 2018 by YN
}

\author{Yusuke~Namekawa}
\affiliation{Faculty of Pure and Applied Sciences,
             University of Tsukuba, Tsukuba, Ibaraki 305-8571, Japan}
\author{Takeshi~Yamazaki}
\affiliation{Faculty of Pure and Applied Sciences,
             University of Tsukuba, Tsukuba, Ibaraki 305-8571, Japan}
\affiliation{Center for Computational Sciences, University of Tsukuba,
	     Tsukuba, Ibaraki 305-8577, Japan}

\pacs{11.15.Ha, % lattice gauge theory
      12.38.Gc  % lattice QCD
}

\begin{abstract}
We evaluate scattering amplitudes at on-shell and half off-shell for $I=2$ S-wave two-pion system
using the Bethe-Salpeter wave function inside the interaction range in the quenched QCD.
The scattering length and effective range are extracted from these scattering amplitudes.
Quark mass dependence of them is investigated with the pion mass ranged in $0.52$--$0.86$~GeV.
We examine consistency between a result by the conventional finite volume method
and our estimate, as well as the phenomenological value.
\end{abstract}

\maketitle

\setcounter{equation}{0}

\section{Introduction}
\label{sec:introduction}

Lattice QCD has contributed to quantitative understanding of hadrons
from the first principle of the strong interaction.
Hadron scattering represented by the scattering length $a_0$, the effective range $r_{\rm eff}$,
and the scattering phase shift $\delta(k)$ itself, can be obtained by lattice QCD.
The recent works are summarized in Ref.~\cite{Briceno:2018bnl}.

The scattering information is extracted by the finite volume method and the extensions,
formulated by M.L\"{u}scher~\cite{Luscher:1986pf,Luscher:1990ux}.
An analytic function describes the relation between two-hadron energies in a finite volume
and the scattering phase shift in the infinite volume.
This relation is derived by use of the Bethe-Salpeter (BS) wave function
outside the interaction range $R$ of two hadrons in quantum field theory~\cite{Lin:2001ek,Aoki:2005uf}.
A method to define a potential between hadrons from the BS wave function is also proposed~\cite{Aoki:2009ji}.

An associated issue between the on-shell scattering amplitude and the BS wave function inside $R$ is
argued in the infinite volume~\cite{Lin:2001ek,Aoki:2005uf,Yamazaki:2017gjl}.
The S-wave two-pion scattering amplitude is defined by the Lehmann-Symanzik-Zimmermann (LSZ) reduction formula
under an assumption that inelastic scattering effects are negligible.
In the center of mass frame, the half off-shell amplitude $H(p;k)$ is related to the pion four-point function,
\begin{eqnarray}
 &&
 e^{-i \vec{q} \cdot \vec{x}} \frac{-i\sqrt{Z}}{-\vec{q}^2 + m_\pi^2 - i \epsilon}
 \frac{4 E_p E_k}{E_p + E_k} H(p;k)
 \nonumber
 \\
 &&
 = \int d^{4}z \, d^{4}y_1 \, d^{4}y_2 \,
 K(\vec{p},\vec{z}) K(-\vec{k}_1,\vec{y}_1) K(-\vec{k}_2,\vec{y}_2)
 \nonumber
 \\
 &&
 \times \langle 0 | \pi_1(\vec{z}) \pi_2(\vec{x}) 
 \pi_1(\vec{y}_1) \pi_2(\vec{y}_2) | 0 \rangle ,
 \label{eq:LSZ}
\end{eqnarray}
where the vectors are four-dimensional.
The two-pion energy is $E_k = 2 \sqrt{m_\pi^2 + k^2}$, where $k = |{\bf k}|$ is a magnitude of three-dimensional momentum.
The kernel is $K(\vec{p},\vec{z}) = i e^{i \vec{p} \cdot \vec{z}} (-\vec{p}^2 + m^2)/{\sqrt{Z}}$
with $Z$ being the renormalization factor of the pion operator $\pi_i$.
The three momenta $\vec{p}, \vec{k_1}$, and $\vec{k_2}$ are on-shell.
Contrarily, $\vec{q}$ is in general off-shell.
The on-shell scattering amplitude $H(k;k)$ is associated with the scattering phase shift $\delta(k)$ through
\begin{eqnarray}
 H(k;k) = \frac{4 \pi}{k} e^{i \delta(k)} \sin \delta(k).
 \label{eq:H_kk_infinite_volume}
\end{eqnarray}
Some ratio is needed to cancel out $e^{i \delta(k)}$,
which can not be measured directly on the lattice.

Based on this issue, we accomplished a simulation to obtain scattering amplitudes from the BS wave function inside $R$
on a finite volume lattice~\cite{Namekawa:2017sxs}.
The simulation utilized the isospin $I=2$ S-wave two-pion in the quenched QCD
with the lattice spacing of $a^{-1} = 1.207$~GeV at the pion mass $m_\pi = 0.86$~GeV.
Using the on-shell amplitude, we numerically confirmed agreement between a result by the finite volume method
and that by our approach.
We also presented lattice QCD can successfully give the half off-shell scattering amplitude.
Although the half off-shell scattering amplitude is not an observable in experiments,
it gives constraints on parameters of effective theories and models of the hadron interaction.
Lattice QCD results of the half off-shell scattering amplitude can be useful
as supplementary data to effective theories and models.

In this paper, we extend our previous simulation~\cite{Namekawa:2017sxs}
to investigate the scattering amplitudes with several pion masses $m_\pi = 0.52$--$0.86$~GeV at the same lattice spacing.
The results are extrapolated to the physical point to examine consistency among the previous result
by the conventional finite size method, the phenomenological estimate, and our result.

This paper is organized as follows.
Sec.~\ref{sec:formulation_lattice} devotes the formulation in the lattice theory.
In Sec.~\ref{sec:set_up}, details of our simulation set up are presented.
Sec.~\ref{sec:result} explains our results at each $m_\pi$, as well as those at the physical point.
Comparison with the previous lattice QCD results and the phenomenological values is also given in this section.
Sec.~\ref{sec:summary} summarizes this paper.
Appendix contains the operator dependence of the scattering amplitude,
formulation of the scattering amplitude using the BS wave function in the momentum space,
and discussion on time dependence of the scattering amplitude.

\section{Formulation on the lattice}
\label{sec:formulation_lattice}

Formulation of the scattering amplitude is explained.
We use the same notation in our previous work~\cite{Namekawa:2017sxs},
following Refs.~\cite{Lin:2001ek,Aoki:2005uf,Yamazaki:2017gjl}.
We restrict ourselves to the S-wave scattering of $I=2$ two-pion in the center of mass frame,
assuming inelastic scattering effects are negligible.
The two-pion BS wave function on the lattice $\phi({\bf x};k)$ is obtained by
\begin{equation}
 \phi({\bf x};k) = \langle 0 | \Phi({\bf x},0) | \pi^+ \pi^+, E_k \rangle ,
 \label{eq:def_BS_wave_func}
\end{equation}
where $| \pi^+ \pi^+, E_k \rangle$ is a ground state of two pions,
and $E_k = 2 \sqrt{m_\pi^2 + k^2}$ is the two-pion energy.
$\Phi({\bf x},t)$ is an operator of two pions,
\begin{equation}
 \Phi({\bf x},t) = \sum_{{\bf r}} \pi^+(R_{A_1^+}[{\bf x}] + {\bf r}, t) \pi^+({\bf r}, t) 
 \label{eq:def_pi_pi_operator}
\end{equation}
with $A_1^+$ projection $R_{A_1^+}[{\bf x}]$.
$R_{A_1^+}[{\bf x}]$ is performed to obtain the S-wave scattering state with an assumption
that higher angular momentum contributions of $l \ge 4$ scattering are negligible.
The pion interpolating operator is defined by
\begin{equation}
 \pi^{+}({\bf x},t) = \bar{d}({\bf x},t) \gamma_5 u({\bf x},t).
 \label{eq:def_pi_operator}
\end{equation}
$\phi({\bf x};k)$ is related to a pion four-point 
function $C_{\pi\pi}({\bf x},t)$ such that
\begin{eqnarray}
 C_{\pi\pi}({\bf x},t) &=&
 \langle 0 | \Phi({\bf x},t_{\rm sink}) \Omega^\dagger(t_{\rm src}) | 0 \rangle
 \label{eq:4point_func}
 \\
 &=& C_k \phi({\bf x};k) e^{-E_k t} + \cdots,
 \label{eq:C_k}
\end{eqnarray}
where $\Omega(t_{\rm src})$ is an operator of two-pion at the source time slice $t_{\rm src}$,
and $t = |t_{\rm sink} - t_{\rm src}|$.
The dot term corresponds to excited state contributions.
$C_k$ is an overall constant.

An essential quantity to calculate the scattering amplitude from $\phi({\bf x};k)$
is the reduced BS wave function $h({\bf x};k)$~\cite{Aoki:2005uf,Yamazaki:2017gjl}.
It is defined through $\phi({\bf x};k)$ as
\begin{equation}
 h({\bf x};k) = (\Delta + k^2) \phi({\bf x};k),
 \label{eq:def_hxk_lat}
\end{equation}
where
\begin{equation}
 \Delta f({\bf x}) = \sum_{i=1}^{3} ( f({\bf x}+\hat{i}) + f({\bf x} - \hat{i}) - 2 f({\bf x}) ).
 \label{eq:lat_laplacian}
\end{equation}
$h({\bf x};k)$ possesses an important property that $h({\bf x};k)$ equals to zero
outside the interaction range of two pions $R$, except for the exponential tail,
\begin{equation}
 h({\bf x};k) = 0 \mbox{\ \ \ for } x > R.
 \label{eq:condition_h}
\end{equation}

$h({\bf x};k)$ defines the half off-shell amplitude on the lattice $H_L(p;k)$ with an off-shell momentum $p$,
\begin{equation}
 H_L(p;k) = -\sum_{{\bf x} \in L^3} C_k h({\bf x};k) j_0(p x),
 \label{eq:H_L_pk_lattice}
\end{equation}
where $j_0(p x)$ is the spherical Bessel function.
If $R$ is less than half of the lattice extent $L$, $R < L/2$, and the exponential tail is negligible,
then the range of the summation can be changed from $L$ to $\infty$ due to Eq.~(\ref{eq:condition_h}),
as discussed in Ref.~\cite{Namekawa:2017sxs}.
$H_L(p;k)$ becomes the half off-shell amplitude in the infinite volume $H(p;k)$ in Eq.~(\ref{eq:LSZ}) as
\begin{equation}
 H(p;k) = \frac{H_L(p;k)}{C_{00}},
 \label{eq:H_pk}
\end{equation}
except for an overall finite volume correction, $C_{00} = C_k / F(k,L)$.
$F(k,L)$ is a finite volume correction of the two-pion state,
called the Lellouch and L\"{u}scher factor~\cite{Lellouch:2000pv}.

As explained above, $R < L / 2$ with a negligible exponential tail of $h({\bf x};k)$
is the sufficient condition for the scattering amplitude calculation on the lattice.
We numerically confirm this condition is satisfied, in the later section.

\section{Set up}
\label{sec:set_up}

Our simulation set up is presented.
We perform a quenched QCD simulation following Refs.~\cite{Aoki:2005uf,AliKhan:2001tx}.
We generate 200 gauge configurations on $24^3 \times 96$ lattice by Hybrid Monte Carlo algorithm,
stored at every 100 trajectories.
Our gauge action is Iwasaki-type~\cite{Iwasaki:2011jk, *Iwasaki:1985we}.
The lattice spacing is $a^{-1} = 1.207$~GeV at the bare coupling is $\beta = 2.334$.
Our valence quark action is Clover-type~\cite{Sheikholeslami:1985ij}.
The Clover coefficient is mean field improved, $C_{\rm SW} = 1.398$~\cite{AliKhan:2001tx}.
Our pion masses are in range of $m_\pi = 0.86$--$0.52$~GeV
with the valence quark hopping parameters of $\kappa_{\rm val} = 0.1340$--$0.1369$.
Our simulation parameters are collected in Table~\ref{tab:simulation}.

We also generate gauge configurations on $24^3 \times 64$ with the same setup
to investigate the source operator dependence of the scattering amplitude on the lattice.
Details of the investigation are explained in Appendix~\ref{app:operator-dep}.

The two-pion four-point function $C_{\pi\pi}({\bf x},t)$ in Eq.~(\ref{eq:4point_func}) is calculated
using complex random $Z2$ sources to avoid Fierz contamination,
\begin{equation}
 \Omega(t)
 = \frac{1}{N_r (N_r - 1)}
   \sum_{\substack{i,j=1 \\ i \ne j}}^{N_r}
   \pi^+(t, \eta_i) \pi^+(t, \eta_j),
\end{equation}
where $\Omega(t)$ is the source operator in Eq.~(\ref{eq:4point_func}).
$\pi^+(t,\eta)$ is defined by
\begin{equation}
 \pi^+(t,\eta)
 = \left[\sum_{{\bf x}_1}\overline{d}({\bf x}_1,t) \eta^\dagger({\bf x}_1)\right]
   \gamma_5
   \left[\sum_{{\bf x}_2} u({\bf x}_2,t) \eta({\bf x}_2)\right] .
\end{equation}
$N_r$ is the number of the random sources $\eta_i({\bf x})$,
satisfying the following condition,
\begin{equation}
 \frac{1}{N_r} \sum_{i=1}^{N_r} \eta_i^\dagger({\bf x}) \eta_i({\bf y})
 \xrightarrow[N_r \to \infty]{}
 \delta({\bf x} - {\bf y}).
\end{equation}
$N_r = 4$ is employed in our simulation.
The source resides at a time $t_{\rm src}$ in Eq.~(\ref{eq:4point_func}) and all spatial points,
as well as all colors and spins.
The latter reduces the simulation cost~\cite{Boyle:2008rh}.
In our set up, we found the gain is roughly a factor of three.
We also calculate a single pion correlator $C_\pi(t)$ with the same source
to measure $m_\pi$ in the large time region, $t \gg 1$,
\begin{equation}
 C_\pi(t)
 = \frac{1}{N_r}\sum_{i=1}^{N_r}\sum_{\bf x}
   \langle 0 | \pi^{+}({\bf x},t_{\rm sink}) (\pi^+(t_{\rm src},\eta_i))^\dagger | 0
   \rangle.
\end{equation}

We perform the measurements with every four time slice per configuration,
{\it i.e.}, the total number of $t_{\rm src}$ is 24.
We adopted the periodic boundary condition in space and the Dirichlet boundary condition in time.
Distance between the Dirichlet boundary and $t_{\rm src}$ is kept to be 12.

\begin{table}[tb]
\begin{center}
\begin{tabular}{cccc}
 \hline \hline
 Lattice size     &  $\kappa_{\rm val}$    &  $N_{\rm src}$  &  $N_{\rm config}$
 \\
 \hline
 $24^3 \times 96$ &  0.1340,0.1358,0.1369  &  24             &  200
 \\
 \hline \hline
\end{tabular}
 \caption{
  \label{tab:simulation}
  Simulation parameters.
 }
\end{center}
\end{table}

We employ two methods to determine the interaction momentum $k$.
One is a momentum from $E_k$, denoted by $k_t$,
\begin{eqnarray}
 k_t^2 = \frac{E_k^2}{4} - m_\pi^2.
 \label{eq:kt}
\end{eqnarray}
$E_k$ is obtained from the temporal correlator of two pions
using $C_{\pi\pi}({\bf x},t)$ in Eq.~(\ref{eq:4point_func}),
\begin{equation}
 C_{\pi\pi}(t) = \sum_{\bf x} C_{\pi\pi}({\bf x},t).
\end{equation}
The other is a momentum from the BS wave function $\phi({\bf x};k)$ outside the interaction range, denoted by $k_s$,
\begin{eqnarray}
 k_s^2 = - \frac{\Delta \phi({\bf x};k)}
                       {\phi({\bf x};k)},
 \mbox{\ \ \ } x > R.
 \label{eq:ks}
\end{eqnarray}
The condition Eq.~(\ref{eq:condition_h}) is convinced by definition.

\section{Result}
\label{sec:result}

\subsection{Effective mass and energy}

\begin{figure*}[!t]
 \centering
 \includegraphics[scale=0.60]{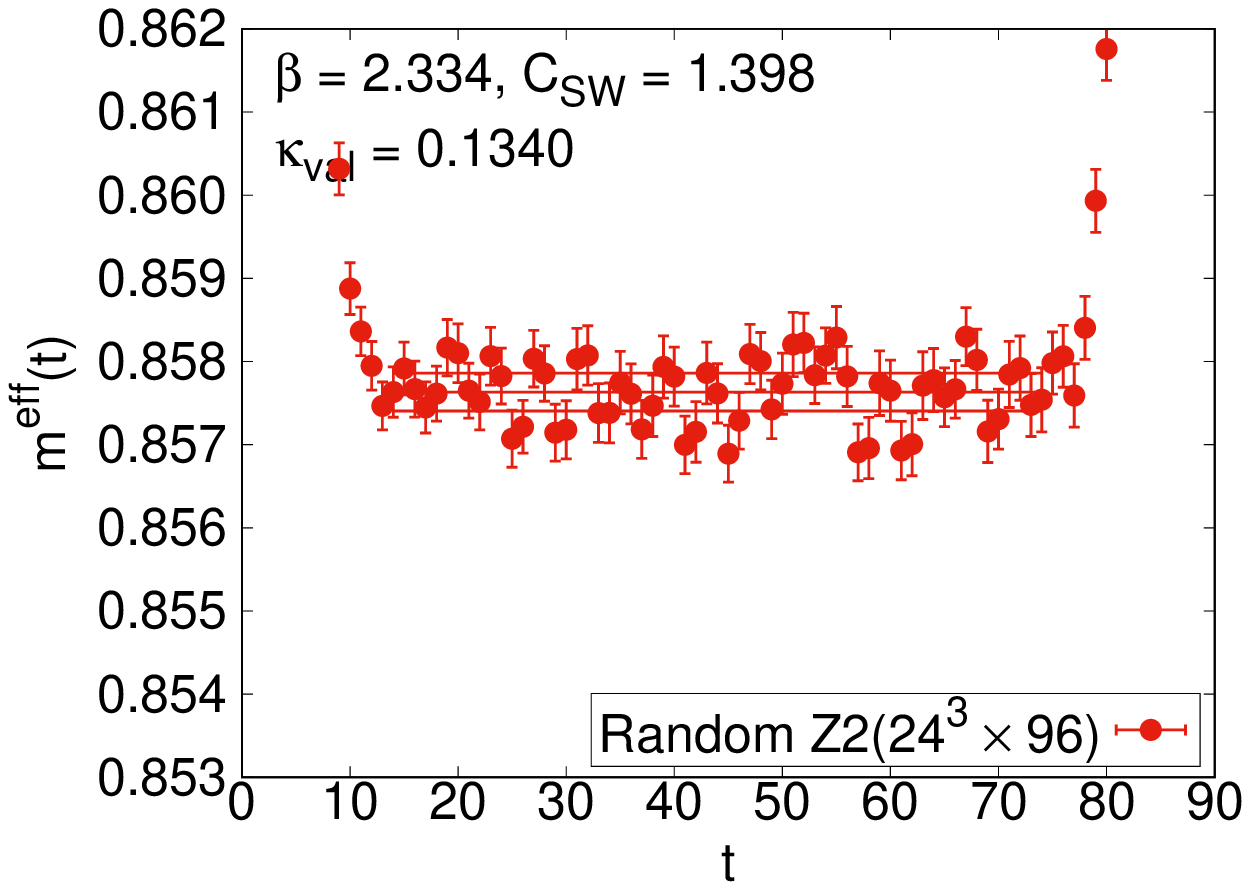}
 \includegraphics[scale=0.60]{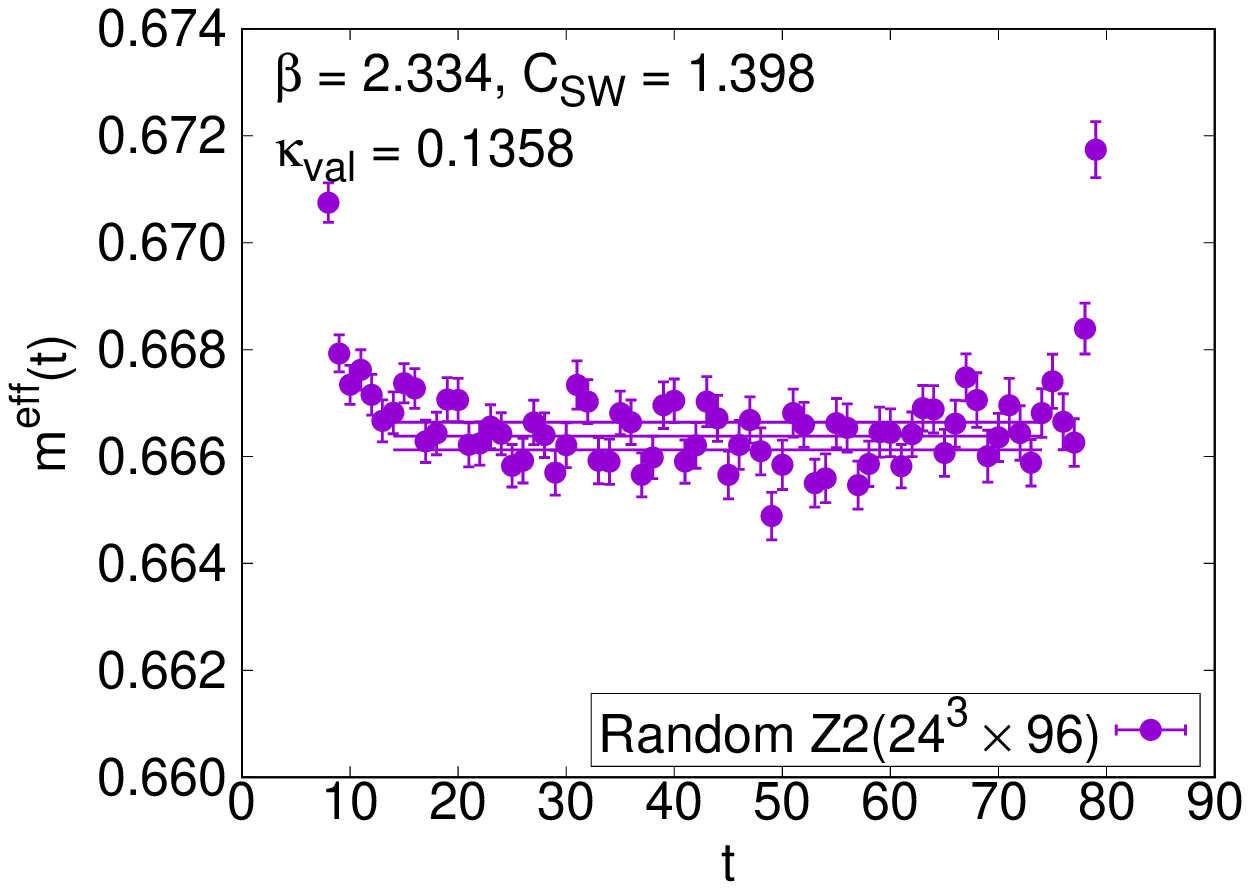}
 \includegraphics[scale=0.60]{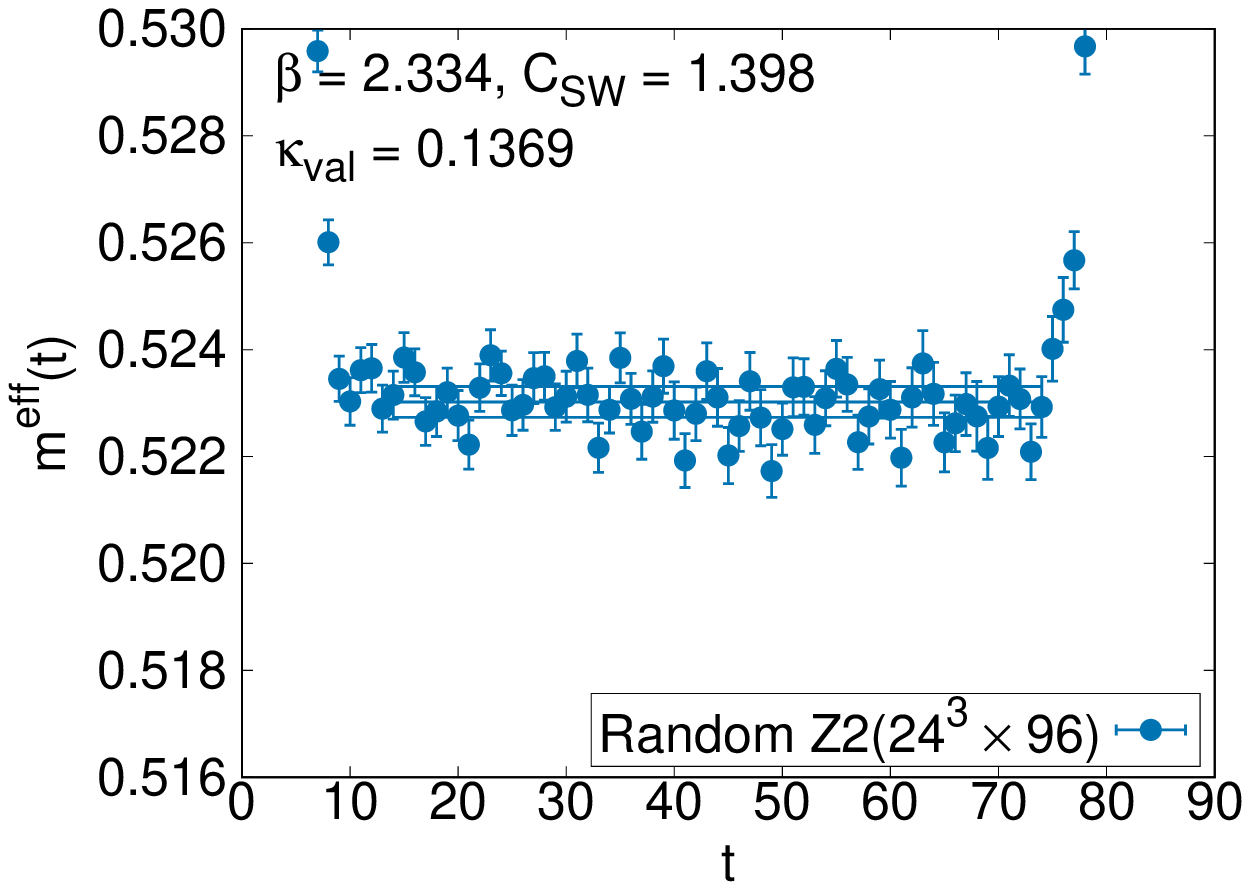}
 \caption{
  \label{fig:m_eff}
  Effective masses of a single pion.
  Results of a single exponential fit with $1 \sigma$ error are denoted by the solid lines.
 }
\end{figure*}

Figure~\ref{fig:m_eff} represents our results of effective masses of a single pion,
defined by
\begin{eqnarray}
 m^{\rm eff}(t) = \log \left( \frac{C_\pi(t)}{C_\pi(t + 1)} \right).
 \label{eq:m_eff}
\end{eqnarray}
A plateau of the effective mass starts from $t=14$ in all $\kappa_{\rm val}$ cases.
We determine $m_\pi$ from a single exponential fit to $C_\pi(t)$
in the range of $[t_{\rm min},t_{\rm max}] = [14,74]$.
The values of $m_\pi$ are listed in Table~\ref{tab:m_pi_E_k_24x96}.

\begin{table}[t]
\begin{center}
\begin{tabular}{ccc}
 \hline \hline
 $\kappa_{\rm val}$  &  $m_\pi$[GeV]  &  $E_k$[GeV]
 \\ \hline
 0.1340              &  0.85763(23)   &  1.71703(47)
 \\
 0.1358              &  0.66638(26)   &  1.33514(52)
 \\
 0.1369              &  0.52302(29)   &  1.04889(60)
 \\
 \hline \hline
\end{tabular}
 \caption{
  \label{tab:m_pi_E_k_24x96}
  $m_\pi$ and $E_k$ on $24^3 \times 96$.
 }
\end{center}
\end{table}

\begin{figure*}[!t]
 \centering
 \includegraphics[scale=0.60]{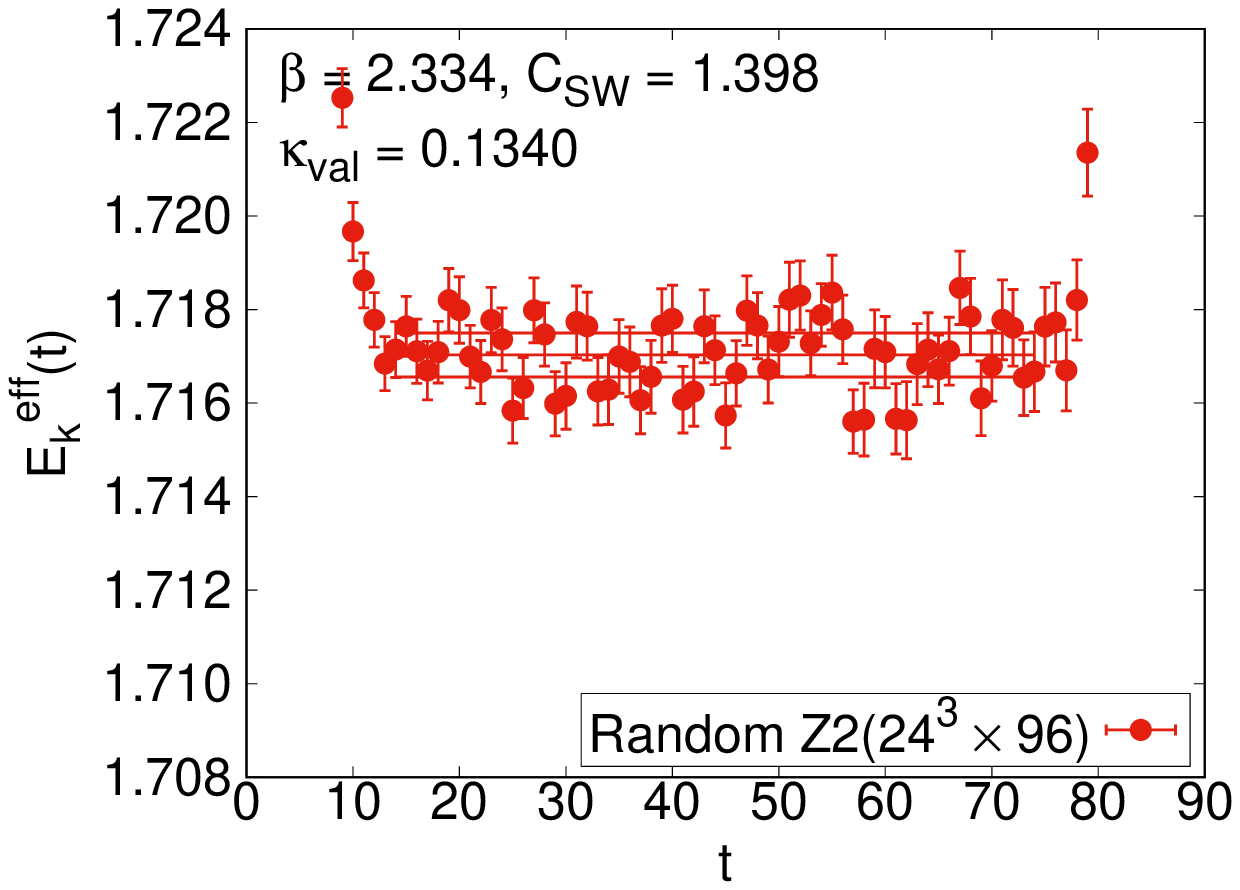}
 \includegraphics[scale=0.60]{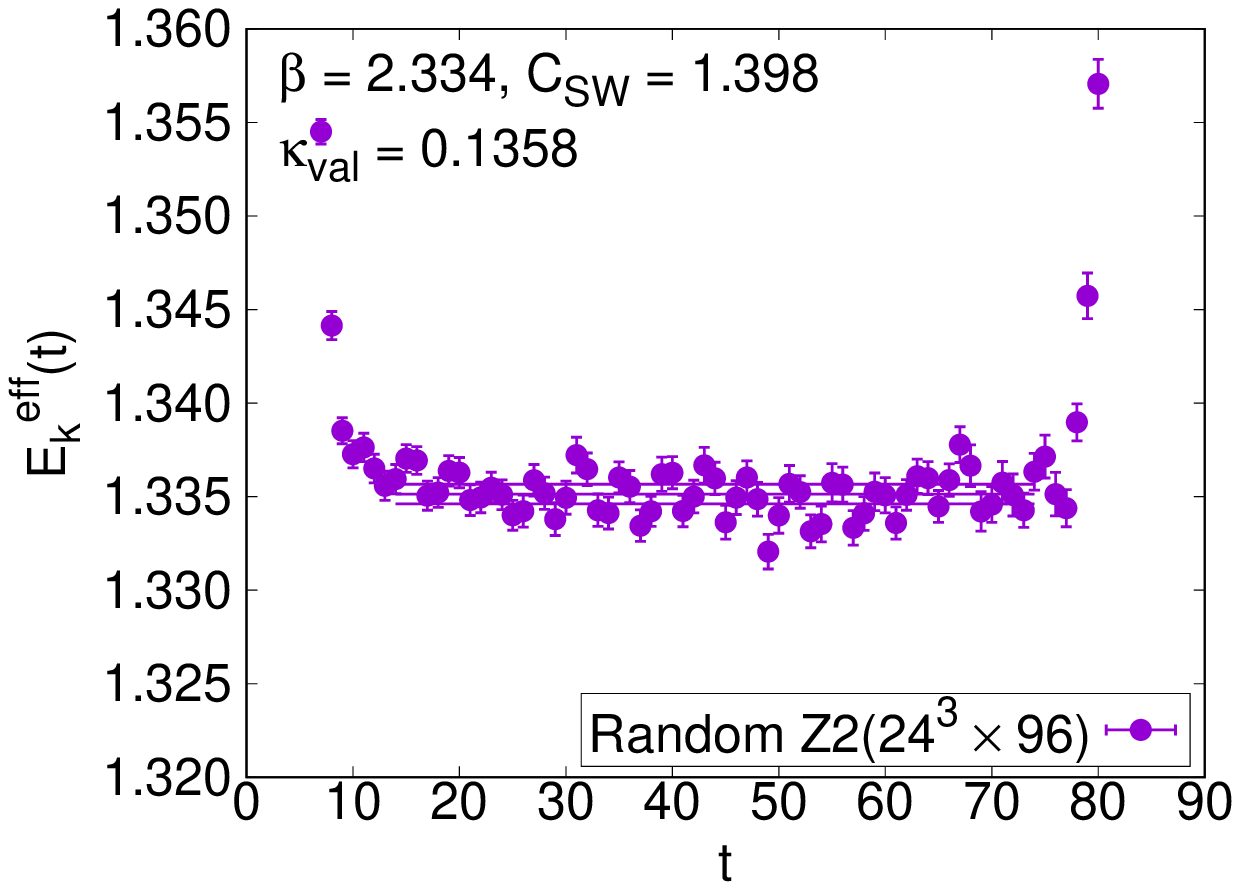}
 \includegraphics[scale=0.60]{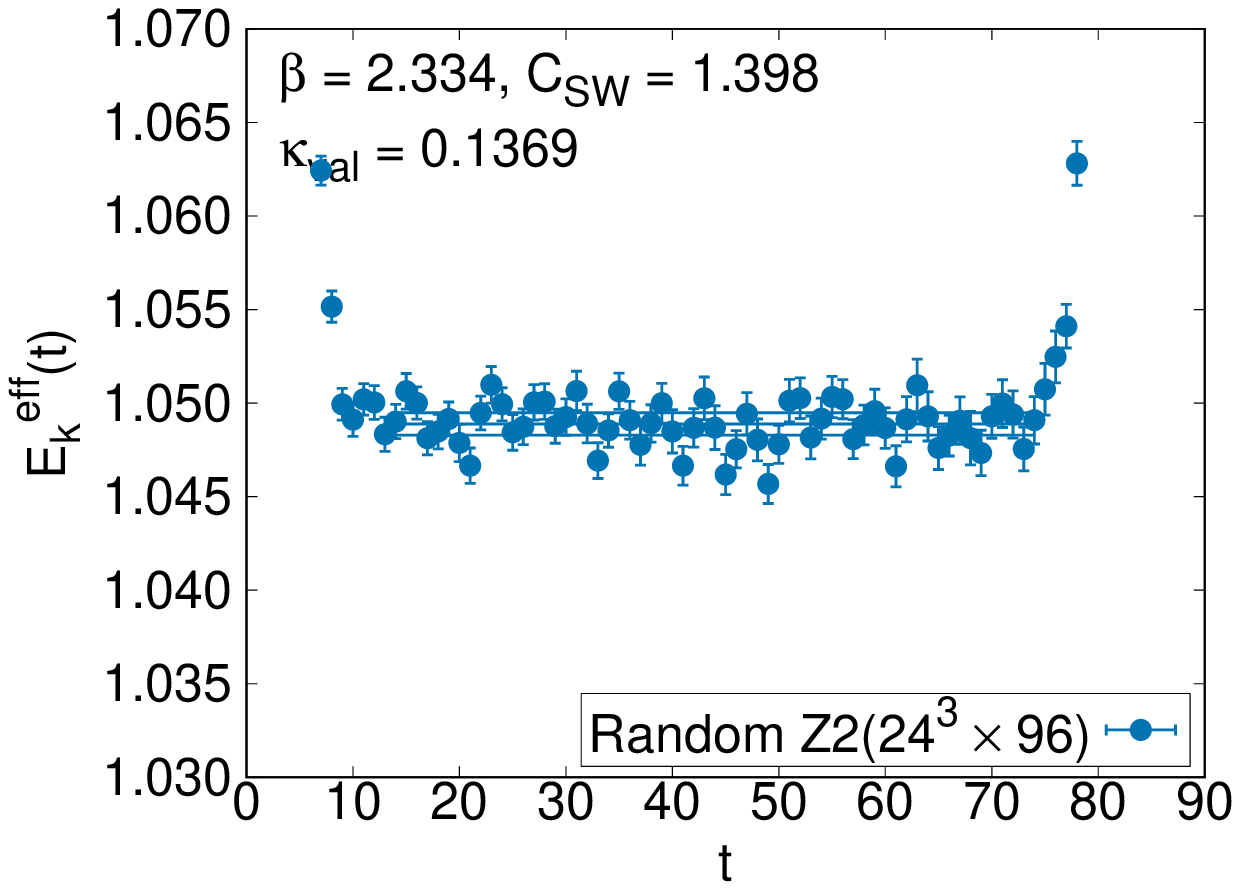}
 \caption{
  \label{fig:E_k_eff}
  Effective energies of $I=2$ two pions.
  Results of a single exponential fit with $1 \sigma$ error are denoted by the solid lines.
 }
\end{figure*}

Figure~\ref{fig:E_k_eff} plots effective energies of $I=2$ two pions defined by
\begin{eqnarray}
 E_k^{\rm eff}(t) = \log \left( \frac{C_{\pi\pi}(t)}{C_{\pi\pi}(t + 1)} \right).
 \label{eq:E_eff}
\end{eqnarray}
We employ the same temporal fitting procedure as that of a single pion
to determine an interaction energy of two pions.
The fitted values of $E_k$ is summarized in Table~\ref{tab:m_pi_E_k_24x96}.

\begin{table}[t]
\begin{center}
\begin{tabular}{ccc}
 \hline \hline
 $\kappa_{\rm val}$  &  $k_t^2$[GeV$^2$]            &  $k_s^2$[GeV$^2$]
 \\ \hline
 0.1340              &  1.513(54) $\times 10^{-3}$  &  1.549(20) $\times 10^{-3}$
 \\
 0.1358              &  1.582(48) $\times 10^{-3}$  &  1.519(19) $\times 10^{-3}$
 \\
 0.1369              &  1.488(48) $\times 10^{-3}$  &  1.497(23) $\times 10^{-3}$
 \\
 \hline \hline 
\end{tabular}
 \caption{
  \label{tab:k2_24x96}
  $k_t^2$ and $k_s^2$ on $24^3 \times 96$.
 }
\end{center}
\end{table}

Using the results for $m_\pi$ and $E_k$ obtained from the fits,
$k_t^2$ in Eq.~(\ref{eq:kt}) is evaluated.
The values of $k_t^2$ are tabulated in Table~\ref{tab:k2_24x96}.

\subsection{BS wave function}
\label{sec:BSwf}

\begin{figure*}[t]
 \centering
 \includegraphics[scale=0.60]{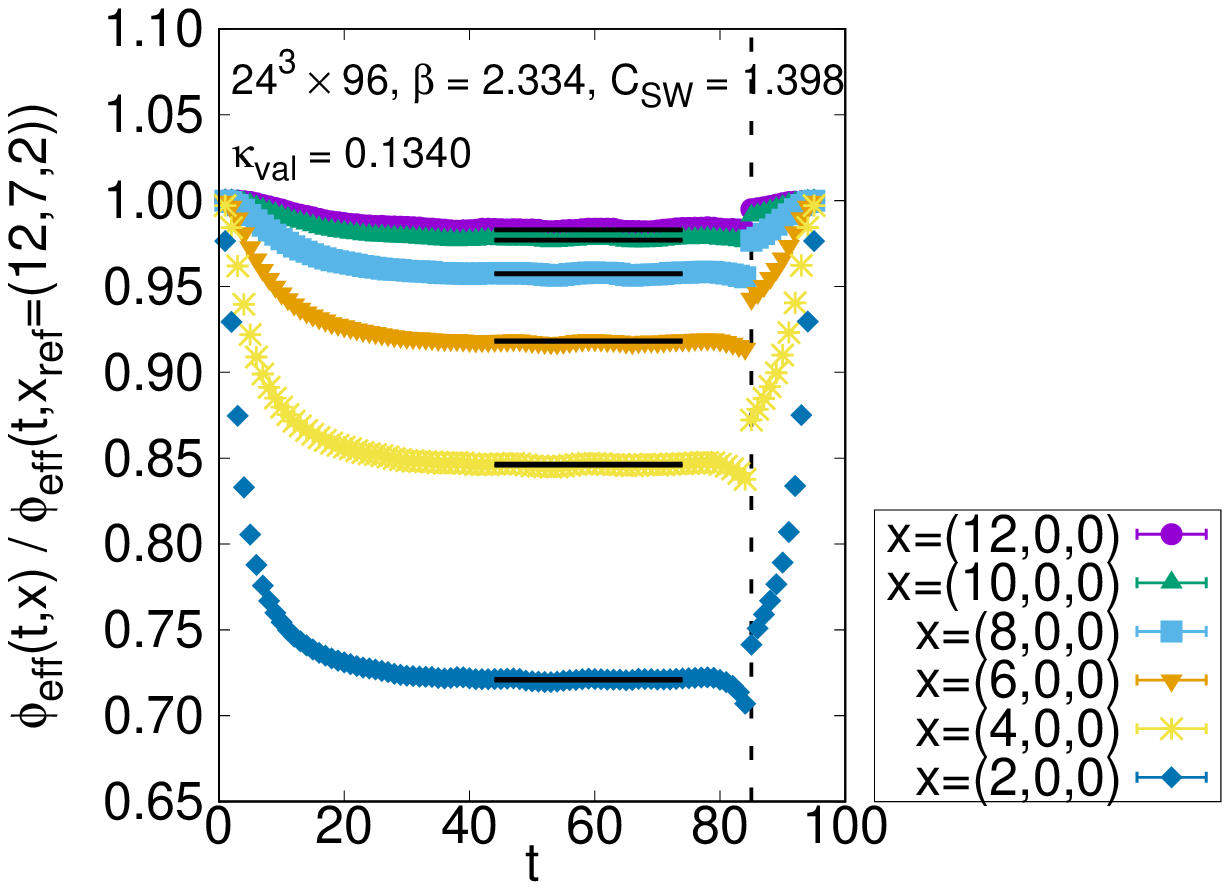}
 \includegraphics[scale=0.60]{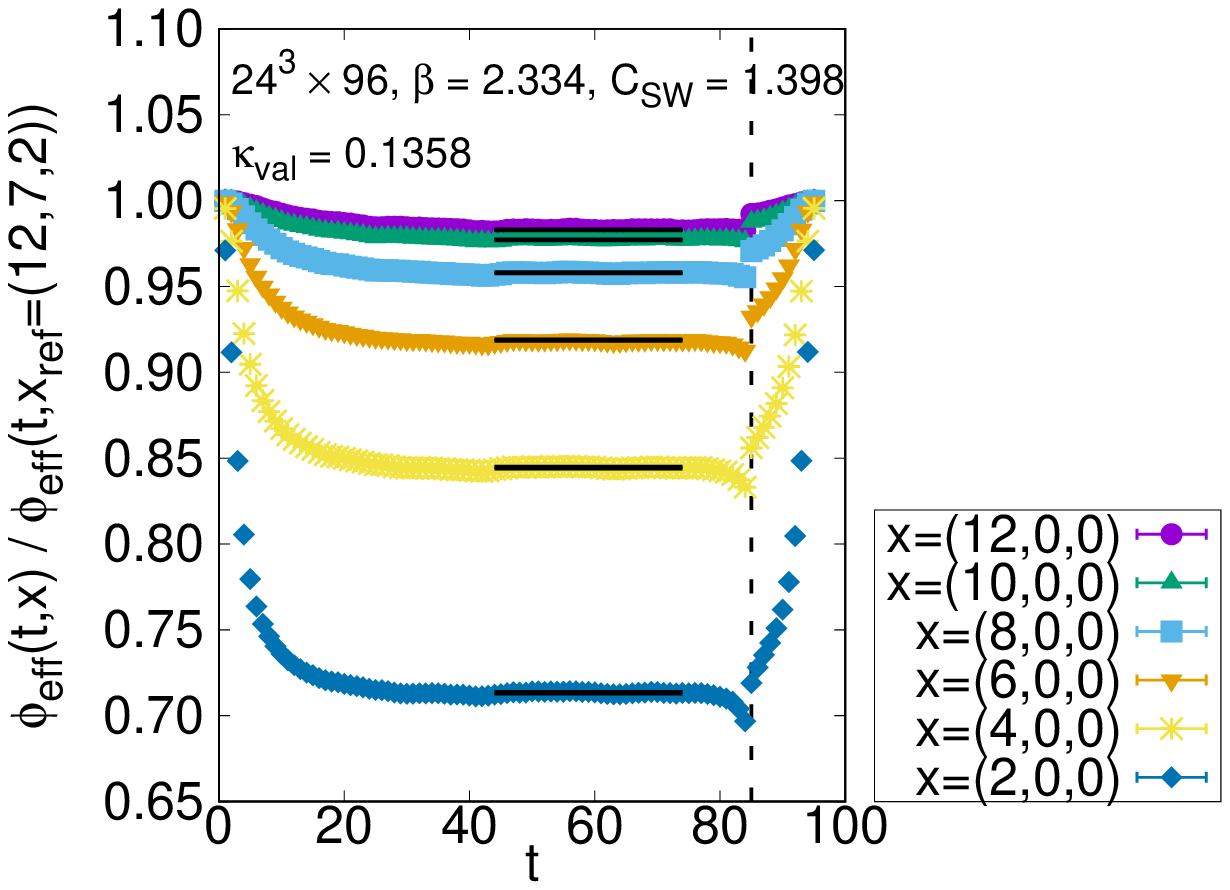}
 \includegraphics[scale=0.60]{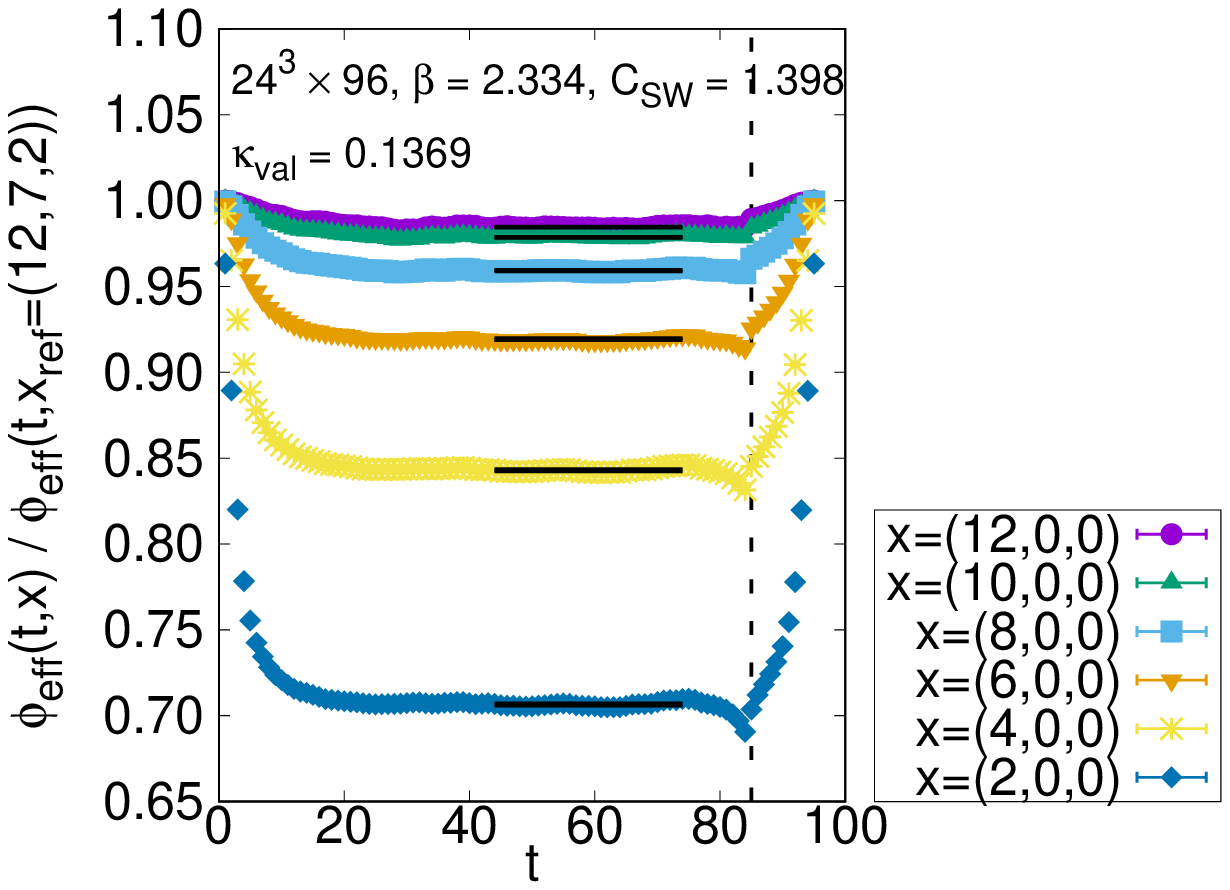}
 \caption{
  \label{fig:wavefunc_eff_24x96}
  Ratios of effective BS wave functions of two pions with the fit results represented by solid horizontal lines.
  The vertical dotted line denotes the Dirichlet boundary position.
 }
\end{figure*}

The BS wave function $\phi({\bf x};k)$ in Eq.~(\ref{eq:def_BS_wave_func}) is calculated as follows.
An effective BS wave function is defined by
\begin{eqnarray}
\frac{\phi_{\rm eff}(t,{\bf x})}
     {\phi_{\rm eff}(t,{\bf x}_{\rm ref})}
 =
 \frac{C_{\pi\pi}(t,{\bf x})}
      {C_{\pi\pi}(t,{\bf x}_{\rm ref})}.
\end{eqnarray}
We choose the reference position of ${\bf x}_{\rm ref}=(12,7,2)$.
$\phi_{\rm eff}(t,{\bf x}) / \phi_{\rm eff}(t,{\bf x}_{\rm ref})$ is plotted in Fig.~\ref{fig:wavefunc_eff_24x96}.
$\phi_{\rm eff}(t,{\bf x})/\phi_{\rm eff}(t,{\bf x}_{\rm ref})$ monotonically decreases with $t$ in an early $t$ region,
where excited state contributions are clearly seen.
A longer time separation is needed for the BS wave functions than those for a pion mass and two-pion energy.
Boundary effects are also observed in the large $t$ region near the Dirichlet boundary position.
The plateau of BS wave functions is observed in $t = 44$--$74$.
We extract $\phi({\bf x};k)$ by a constant fit to $\phi_{\rm eff}(t,{\bf x})$ in all ${\bf x}$
combined with the single exponential fit in the range of $[t_{\rm min},t_{\rm max}] = [44,74]$.

\subsection{Sufficient condition}

\begin{figure*}[t]
 \centering
 \includegraphics[scale=0.60]{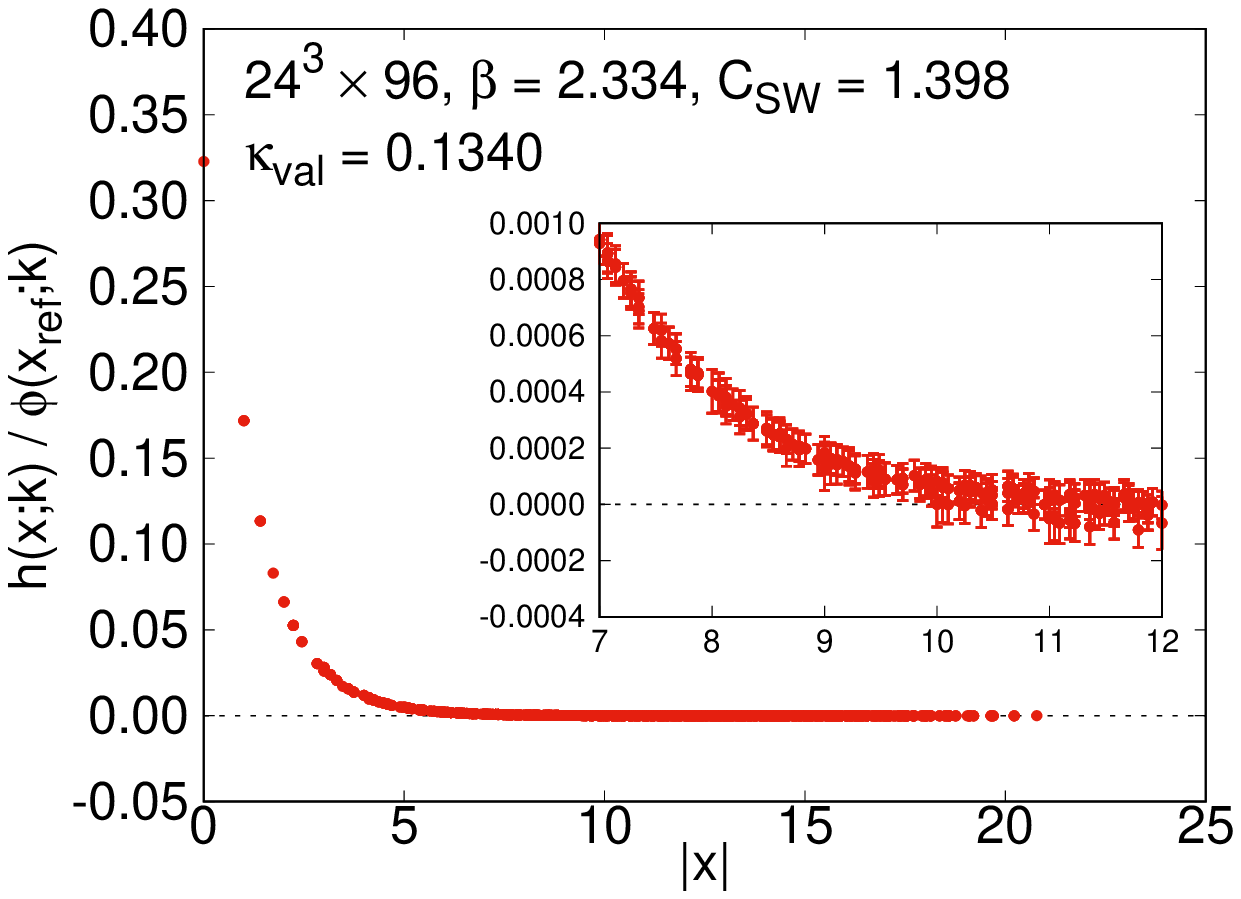}
 \includegraphics[scale=0.60]{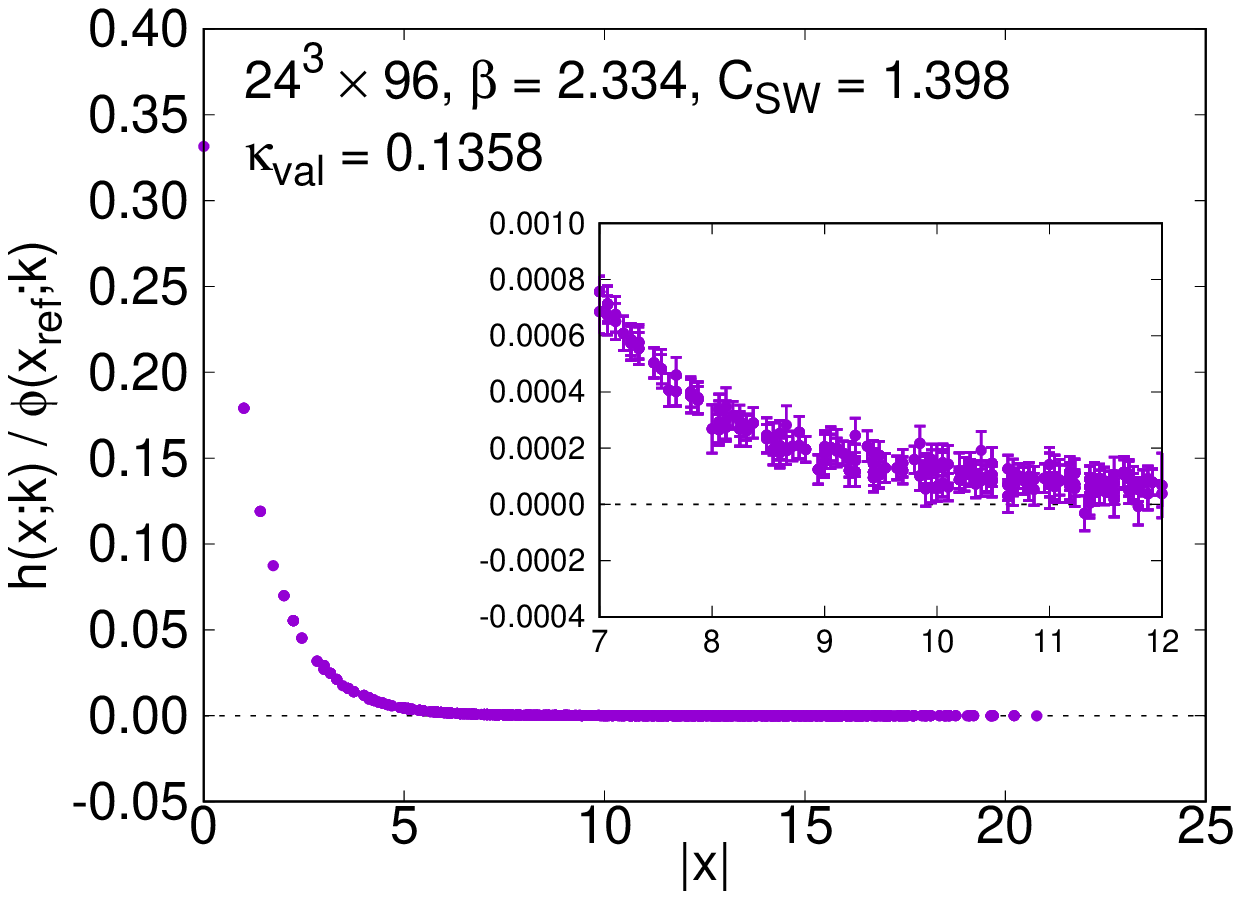}
 \includegraphics[scale=0.60]{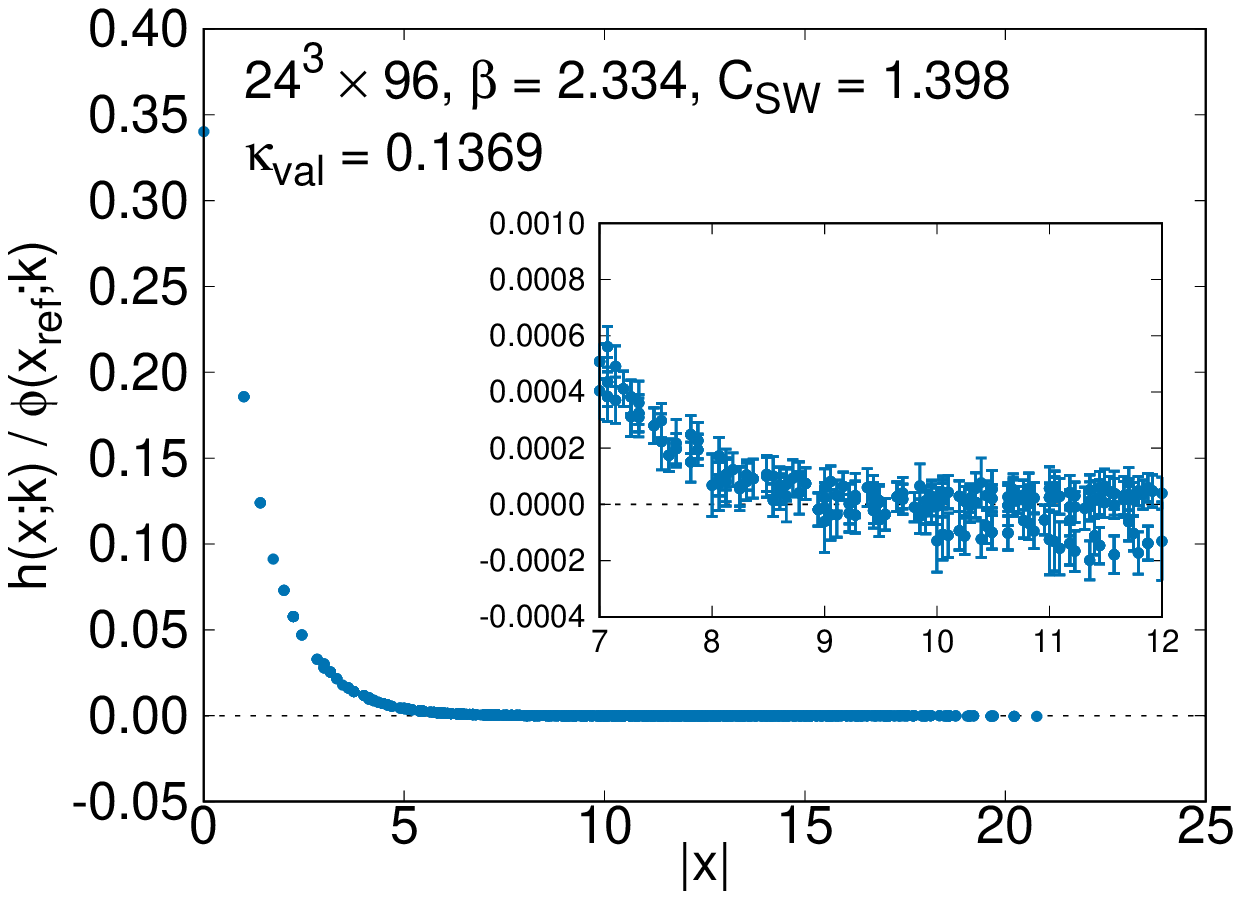}
 \caption{
  \label{fig:reduced_phi_L}
  Ratio of the reduced wave function $h({\bf x};k)$ over the wave function $\phi({\bf x}_{\rm ref};k)$
  at a reference point ${\bf x}_{\rm ref} = (12,7,2)$ using $k = k_t$.
  The inside panel enlarges data in $7 \le x \le 12$.
 }
\end{figure*}

We confirm the sufficient condition of Eq.~(\ref{eq:H_L_pk_lattice}) is satisfied in our simulation.
The reduced BS wave function $h({\bf x};k)$ in Eq.~(\ref{eq:def_hxk_lat}) is calculated
using the fit result of $\phi_{\rm eff}(t,{\bf x})$ and $k_t^2$ tabulated in Table~\ref{tab:k2_24x96}.
Figure~\ref{fig:reduced_phi_L} illustrates our results of $h({\bf x};k)$.
We employ a ratio of $h({\bf x};k)$ over $\phi({\bf x}_{\rm ref};k)$ at a reference point ${\bf x}_{\rm ref}$
to cancel out the overall factor.
For $x \gtrsim 10$, $h({\bf x};k_t) = 0$ is found to be satisfied in our statistical precision.
This result shows the interaction range $R \sim 10 < L / 2$,
and the exponential tail of $h({\bf x};k)$ is negligible compared to our statistical error.
Our data guarantees the sufficient condition of Eq.~(\ref{eq:H_L_pk_lattice}) in our quark mass region.

Using data in the outside region of $R$, an alternative interacting momentum $k_s^2$ in Eq.~(\ref{eq:ks})
can be determined, which is more precise than $k_t^2$~\cite{Aoki:2005uf}.
We obtain $k_s^2$ from a constant fit to $-\Delta \phi({\bf x};k)/\phi({\bf x};k)$
with the fit range of $[x_{\rm min}, x_{\rm max}] = [10, 12 \sqrt{3}]$.
Table~\ref{tab:k2_24x96} collects our results of $k_s^2$, as well as $k_t^2$.
$k_s^2$ is consistent with $k_t^2$ with a smaller error than that of $k_t^2$ by a factor of two.

\subsection{Scattering amplitude}
\label{sec:scattering_amplitude}

Once the sufficient condition $R \sim 10 < L / 2$ is satisfied,
the scattering amplitude can be computed using Eq.~(\ref{eq:H_L_pk_lattice}).
We choose $k = k_s$ in the following analyses, unless explicitly stated.

\begin{figure}[!tb]
 \centering
 \includegraphics[scale=0.60]{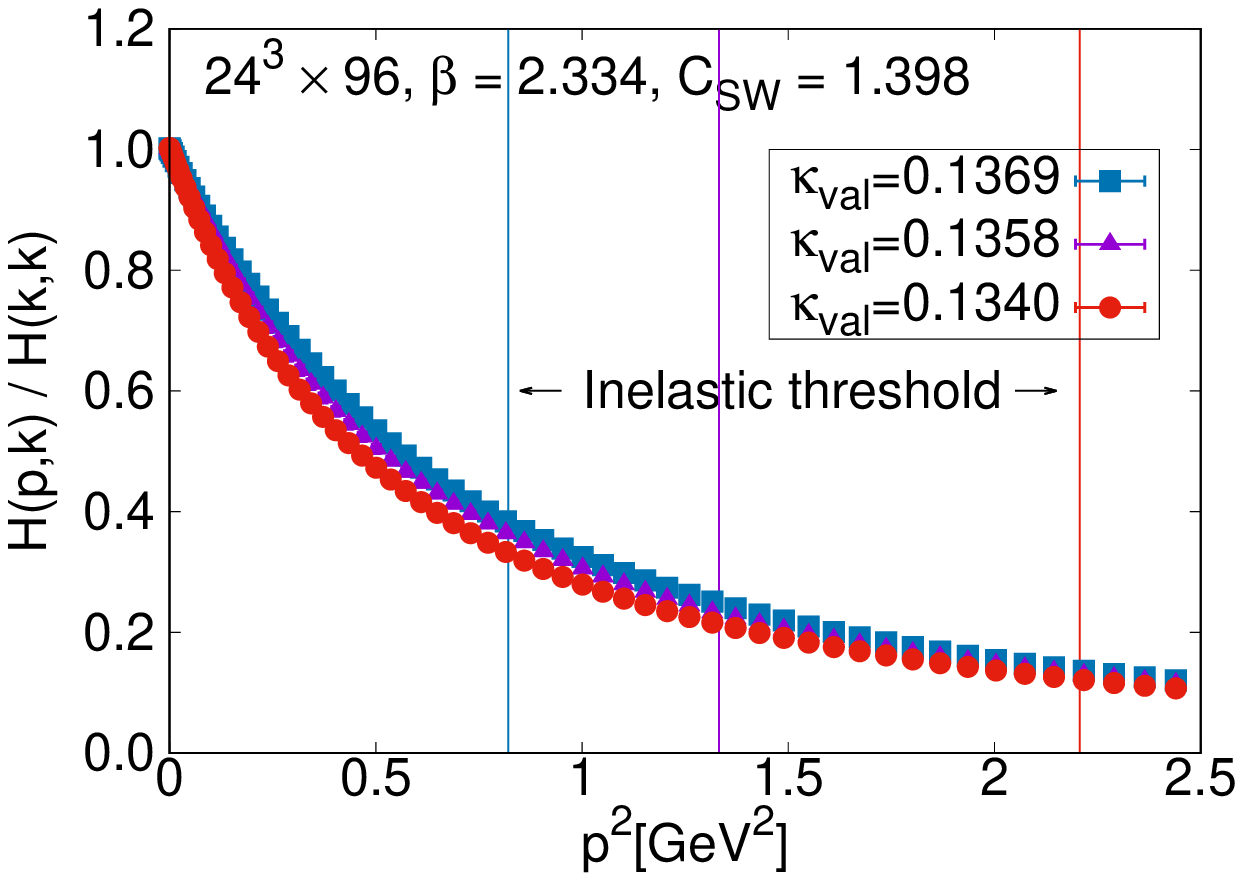}
 \caption{
  \label{fig:half_off_shell}
  Momentum dependence of a ratio of the scattering amplitude at half off-shell $H(p;k)$
  over the value at on-shell $H(k;k)$.
  The vertical line expresses the threshold momentum of the two-pion scattering.
 }
\end{figure}

Figure~\ref{fig:half_off_shell} displays off-shell momentum dependence of the half off-shell amplitude $H(p;k)$
in Eq.~(\ref{eq:H_pk}).
The overall factor of $H_L(p;k)$ in Eq.~(\ref{eq:H_L_pk_lattice}) are eliminated
by taking a ratio of $H_L(p;k)$ over its on-shell value $H_L(k;k)$,
\begin{equation}
 \frac{H(p;k)}{H(k;k)} = \frac{H_L(p;k)}{H_L(k;k)}.
 \label{eq:rat-offshell-onshell}
\end{equation}
A clean signal of the ratio is observed throughout our $p^2$ range.
The validity of $H(p;k)$ is ensured below the threshold
drawn in the figure at $p^2 = 3 m_\pi^2$ i.e. $E_k = 4 m_\pi$,
though the quenched approximation prohibits dynamical inelasticity.

The operator dependence of $H_L(p;k)$ is examined for both source and sink operators.
The dependence is under control in our simulations.
The details are explained in Appendix~\ref{app:operator-dep}.

We discuss the lattice artifacts in our result of $H_L(p;k)$.
The rotational symmetry breaking at the finite lattice spacing causes deviation
between on-axis and off-axis $h({\bf x};k)$ values.
The influence to $H_L(p;k)$ is evaluated to be 3 \% in our simulation at $a^{-1} = 1.207$~GeV.
The size of the error is comparable to our statistical error.
Another issue is the finite lattice artifact in the short distance.
It highly affects the data especially around $x=0$.
They are suppressed in $H_L(p;k)$, however, due to the Jacobian factor $r^2$ in the integration,
\begin{eqnarray}
 H_L(p;k)
 &=& - \sum_{i=1}^{3} \sum_{x_i = -L/2 + 1}^{L/2} C_k h({\bf x};k) j_0(p r)
 \\
 &\simeq& - 4 \pi \int_{0}^{\sqrt{3} L/2} dr \, r^2 C_k h({\bf x};k) j_0(p r).
\end{eqnarray}
Contribution from $h({\bf x};k)$ near $x=0$ is not significant.
Nevertheless, the continuum limit is required to remove these lattice artifacts.

$H_L(p;k)$ can be also calculated with the BS wave function
in the momentum space, $\widetilde{\phi}(p;k) = \sum_{\bf x}\phi({\bf x};k)e^{-i {\bf p}\cdot{\bf x}}$
(see for example Ref.~\cite{Carbonell:2016ekx}).
We numerically confirmed both approaches give a consistent value.
Appendix~\ref{sec:formulation_momentum_space} explains details of the formulation in the momentum space.

\begin{figure}[!t]
 \centering
 \includegraphics[scale=0.60]{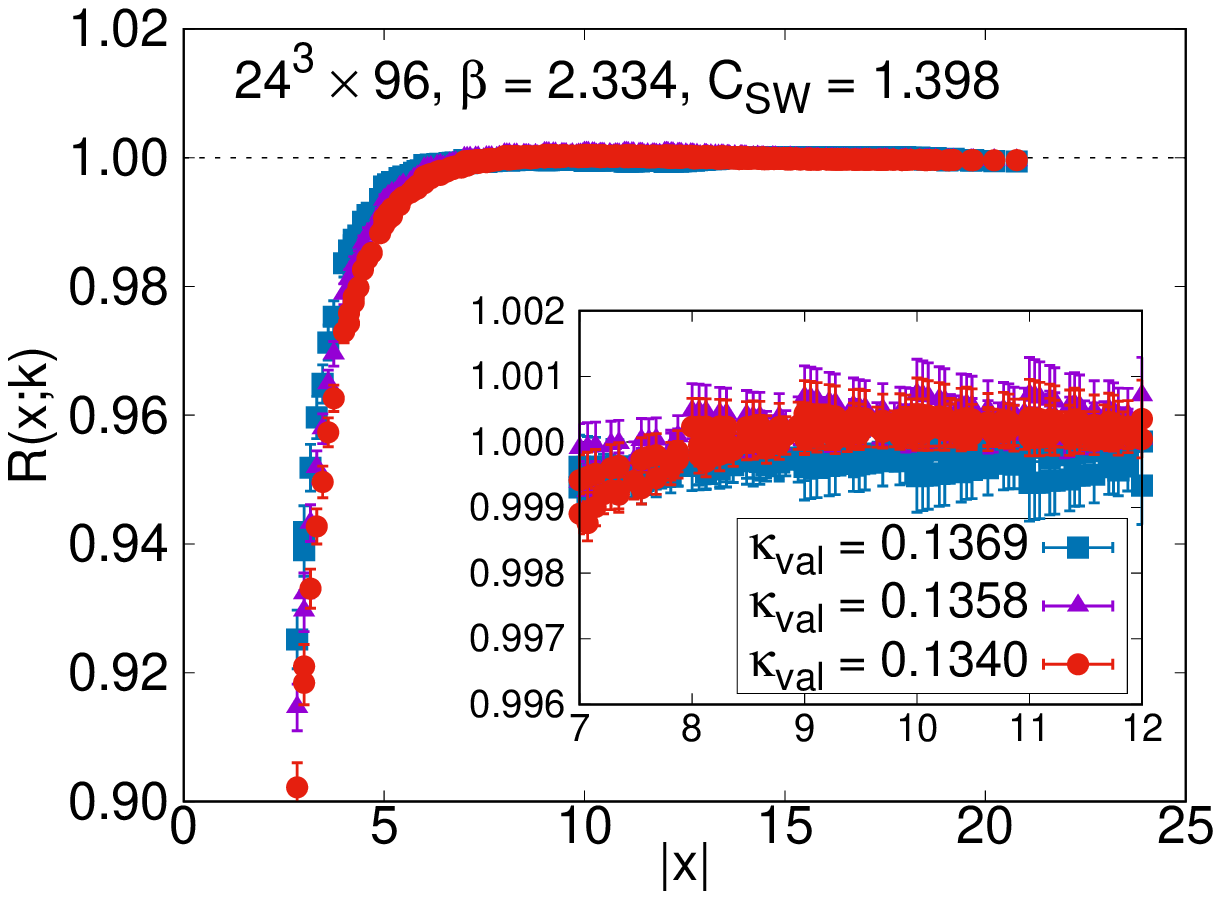}
 \caption{
  \label{fig:test_hkk}
  ${\bf x}$ dependence of the ratio $R({\bf x};k)$ defined in Eq.~(\ref{eq:def_R}).
 }
\end{figure}

Two supplemental confirmations of the validity of $R \sim 10$ are possible by use of the scattering amplitude.
One is the range of the summation for $H_L(p;k)$ in Eq.~(\ref{eq:H_L_pk_lattice}).
$H_L(p;k)$ using a summation over all spatial volume gives a consistent value
with that using a summation up to $x = 10 \sim R$, which implies correctness of the estimate of $R \sim 10$.
The other utilizes the on-shell scattering amplitude $H_L(k;k)$
and an analytic solution of $(\Delta+k^2)\phi({\bf x};k) = 0$ in $x > R$.
In the S-wave case, the analytic solution $\phi_{x > R}({\bf x};k)$ can be expressed
using the Green function on the lattice $G({\bf x};k)$,
\begin{eqnarray}
 C_k \phi_{x > R}({\bf x};k)
 &=& v_{00} G({\bf x};k),
 \label{eq:expand_phi_G}
 \\
 G({\bf x};k)
 &=& \frac{1}{L^3}
     \sum_{{\bf p} \in \Gamma} e^{i {\bf x} \cdot {\bf p} } \frac{1}{p^2 - k^2},
 \label{eq:def_G}
 \\
 \Gamma
 &=& \{ {\bf p} | {\bf p} = \frac{ 2 \pi }{ L } {\bf n}, {\bf n} \in {\bf Z}^3 \},
 \label{eq:def_Gamma}
\end{eqnarray}
where $v_{00}$ is a constant.
$\phi_{x > R}({\bf x};k)$ can be also expressed by the phase shift $\delta(k)$,
\begin{eqnarray}
 C_k \phi_{x > R}({\bf x};k)
 &=& C_{00} e^{i\delta(k)} \frac{ \sin(k x + \delta(k)) }{ kx }
 \nonumber \\
 &+& (l \ge 4),
 \label{eq:expand_phi_sin}
\end{eqnarray}
where $(l \ge 4)$ contains only the spherical Bessel functions $j_l(px)$ of $l \ge 4$.
Comparing Eq.~(\ref{eq:expand_phi_G}) with Eq.~(\ref{eq:expand_phi_sin})
using the expansion by $j_0(k x)$ and $l=0$ spherical Neumann function $n_0(k x)$
leads to two simple equations~\cite{Aoki:2005uf}.
The coefficient of $n_0(k x)$ gives
\begin{eqnarray}
 H_L(k;k) = v_{00} .
\label{eq:finite_volume_1}
\end{eqnarray}
In parallel, the coefficient of $j_0(k x)$ provides
\begin{equation}
 k \cot \delta(k) H_L(k;k) = 4 \pi v_{00} g_{00}(k) ,
 \label{eq:finite_volume_2}
\end{equation}
where 
\begin{equation}
g_{00}(k) = \frac{1}{L^3}\sum_{{\bf p}\in \Gamma}\frac{1}{p^2-k^2} . 
\end{equation}
Using Eqs.~(\ref{eq:finite_volume_1}) and (\ref{eq:finite_volume_2}),
one obtains the finite volume formula~\cite{Luscher:1990ux},
\begin{equation}
 k\cot\delta(k) = 4\pi g_{00}(k) .
\end{equation}
Therefore, Eq.~(\ref{eq:finite_volume_1}) must be satisfied in the finite volume method.
Based on this argument, we define an indicator $R({\bf x};k)$ to test the equality in Eq.~(\ref{eq:finite_volume_1}),
\begin{equation}
 R({\bf x};k)
 = \frac{H_L(k;k)}
        {C_k \phi({\bf x};k)}
   G({\bf x};k) .
 \label{eq:def_R}
\end{equation}
Outside the interaction range $R$, $R({\bf x};k)$ becomes unity, if Eq.~(\ref{eq:finite_volume_1}) is satisfied.
Figure~\ref{fig:test_hkk} represents our result of $R({\bf x};k)$.
It increases monotonically and approaches to unity.
$R({\bf x};k)$ is consistent with unity in $x \gtrsim 10$, as expected.
It validates our estimate of $R \sim 10$.

\begin{figure}[!t]
 \centering
 \includegraphics[scale=0.60]{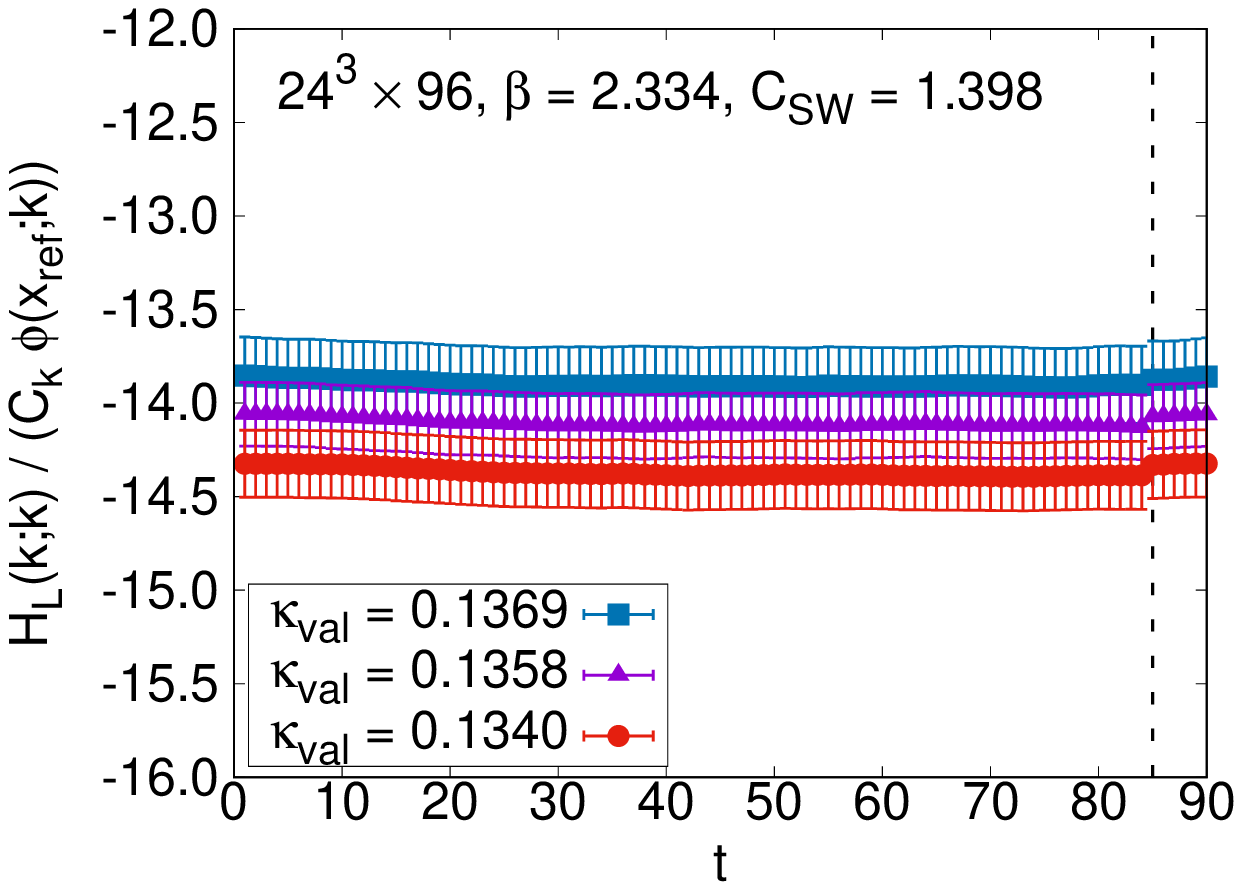}
 \caption{
  \label{fig:t_dependence_of_H_over_phi}
  $t$-dependence of the scattering amplitude over the wave function
  $H_L(k;k) / (C_k \phi({\bf x}_{\rm ref};k))$ with ${\bf x}_{\rm ref} = (12,7,2)$.
  The vertical dotted line denotes the Dirichlet boundary position.
 }
\end{figure}

\begin{figure*}[!t]
 \includegraphics[scale=0.6]{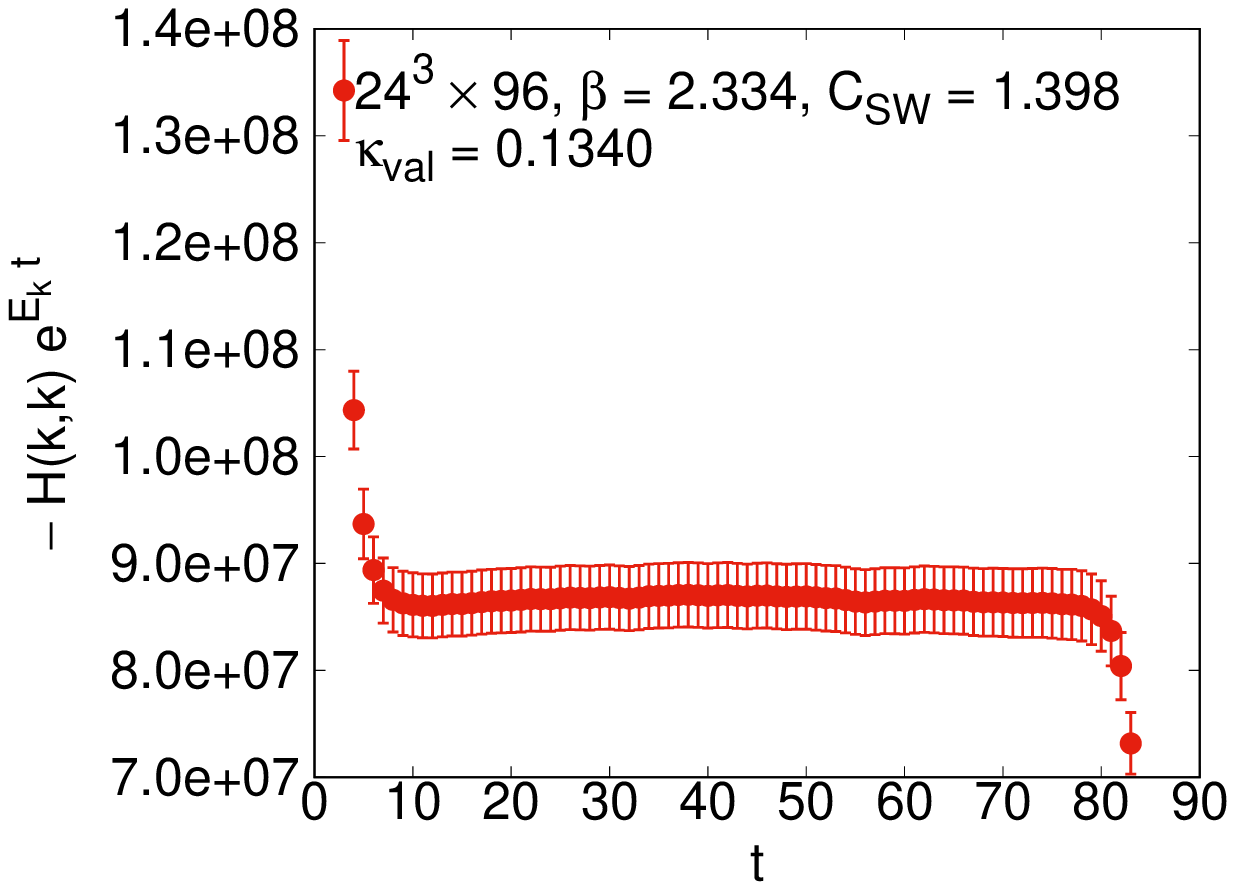}
 \includegraphics[scale=0.6]{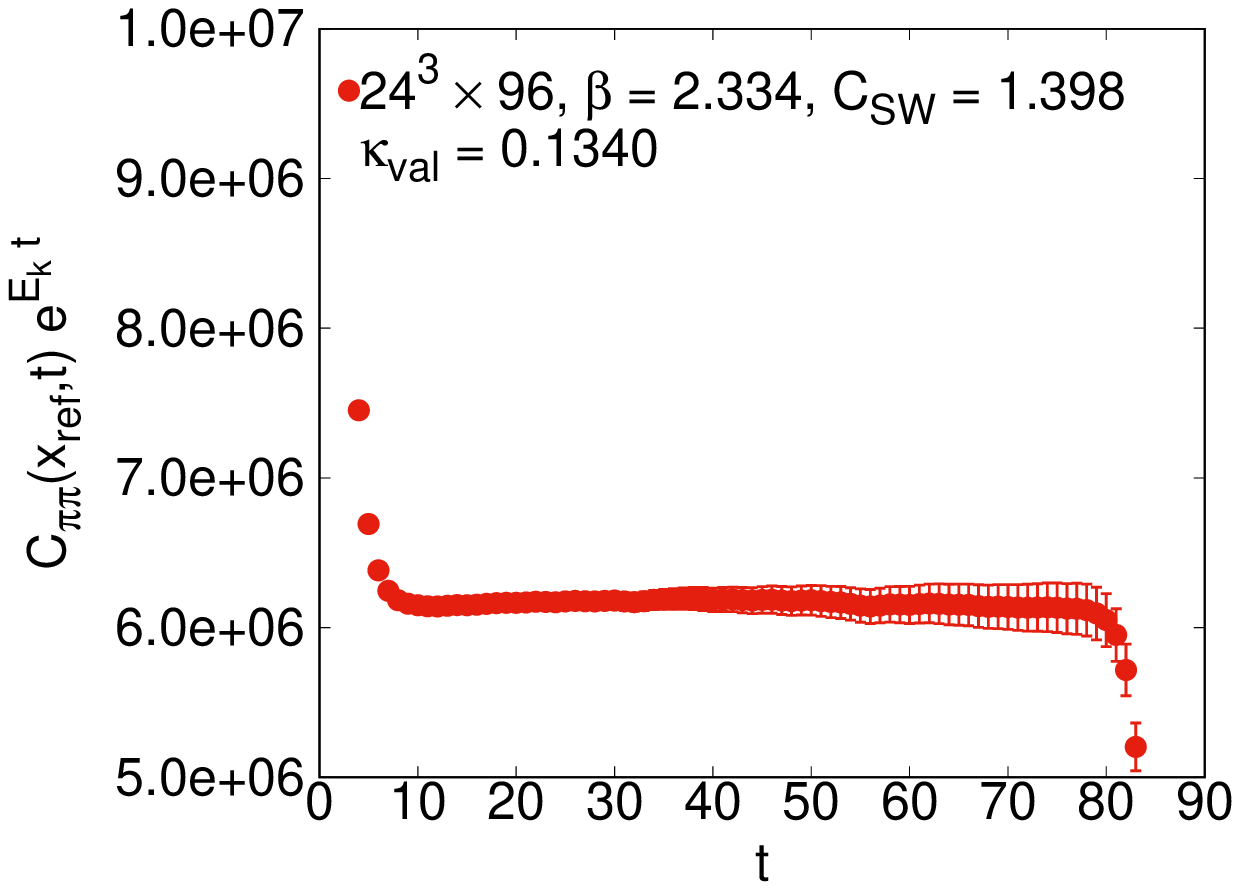}
 \caption{
  \label{fig:each_t_H_phi}
  Numerator and denominator of $H_L(k;k) / (C_k \phi({\bf x}_{\rm ref};k))$
  at $\kappa_{\rm val}=0.1340$ multiplied by $e^{E_k t}$ are plotted in the left and right panels.
 }
\end{figure*}

We also discuss $t$ dependence of the on-shell amplitude $H_L(k;k)$.
$H_L(k;k)$ is calculated at each $t$ for the $t$ dependence analysis,
in contrast to the above calculation using the fit result in Sec.~\ref{sec:BSwf}.
Figure~\ref{fig:t_dependence_of_H_over_phi} illustrates $t$ dependence of $H_L(k;k) / (C_k \phi({\bf x}_{\rm ref};k))$.
The ratio is almost flat in $t$.
Figure~\ref{fig:each_t_H_phi} displays, on the other hand,
the numerator and denominator at $\kappa_{\rm val}=0.1340$ 
multiplied by the exponential factor of the two-pion ground state energy $e^{E_k t}$.
The results show clear excited state contributions in the small $t$ region, $t < 10$.
It should be noticed a choice of $x_{\rm ref}$ varies the excited state contributions of the denominator,
as indicated in Fig.~\ref{fig:wavefunc_eff_24x96}.
Figures~\ref{fig:t_dependence_of_H_over_phi} and \ref{fig:each_t_H_phi} suggest
contributions from the excited states are well compensated in the ratio of $H_L(k;k) / (C_k \phi({\bf x}_{\rm ref};k))$.
The details are discussed in Appendix~\ref{app:ratio}.
Further investigation of the excited state compensation needs the variational method.

\subsection{Physical quantities from scattering amplitudes}

We can extract physical observables from the scattering amplitude.
The scattering phase shift $\delta(k)$ is obtained by the scattering amplitude at on-shell $H_L(k;k)$
and the BS wave function at some reference point outside of the interaction range $x_{\rm ref} > R$,
\begin{eqnarray}
 \frac{H_L(k;k)}{C_k \phi({\bf x}_{\rm ref};k)}
 = \frac{4 \pi x_{\rm ref} \sin \delta(k)}{\sin(k x_{\rm ref} + \delta(k))}.
 \label{eq:H_kk_finite_volume}
\end{eqnarray}
We used the expansion of $C_k \phi({\bf x}_{\rm ref};k)$ in Eq.~(\ref{eq:expand_phi_sin})
and assumed $l \ge 4$ terms are negligible.
The phase factor as well as the overall constants are canceled in the ratio.
Inversely, $\delta(k)$ is given by
\begin{equation}
 \tan \delta(k)
 = \frac{\sin(k x_{\rm ref})}
   {\displaystyle{4 \pi x_{\rm ref} \frac{C_k \phi({\bf x}_{\rm ref};k)}{H_L(k;k)}
 - \cos(k x_{\rm ref})}}.
 \label{eq:tan_delta}
\end{equation}

The reference point is chosen to be ${\bf x}_{\rm ref} = (12,7,2)$ by the following procedure.
Evaluation of $\tan \delta(k)$ through Eq.~(\ref{eq:tan_delta}) requires the $l \ge 4$ terms
in Eq.~(\ref{eq:expand_phi_sin}) must be negligible at ${\bf x}_{\rm ref}$.
We select ${\bf x}_{\rm ref}$ to minimize the leading $l=4$ contribution in the $(l \ge 4)$ terms.
The size of the $l=4$ term is examined using an expansion of $\phi({\bf x};k)$ in $x > R$,
\begin{eqnarray}
 \phi({\bf x};k)
 &=& A_{0}(k) Y_{00}(R_{A_1^+}[{\bf x} / x]) n_0(k x)
 \nonumber \\
 &+& B_{0}(k) Y_{00}(R_{A_1^+}[{\bf x} / x]) j_0(k x)
 \nonumber \\
 &+& B_{4}(k) Y_{40}(R_{A_1^+}[{\bf x} / x]) j_4(k x)
 \nonumber \\
 &+& ( l \ge 6 ),
\end{eqnarray}
where $A_l(k),B_l(k)$ are constants.
$Y_{lm}({\bf x} / x)$ is the spherical harmonic function with $A_1^+$ projector, $R_{A_1^+}$.
It is an alternative expression of Eq.~(\ref{eq:expand_phi_sin}).
Assuming $A_l(k),B_l(k) = O(1)$,
the size of the $l=4$ contribution at each ${\bf x}$ is estimated by using a ratio $Y({\bf x};k)$,
\begin{equation}
 Y({\bf x};k)
 = \frac{Y_{40}(R_{A_1^+}[{\bf x} / x]) j_4(k x)}
        {Y_{00}(R_{A_1^+}[{\bf x} / x]) j_0(k x)} .
 \label{eq:def_Y}
\end{equation}
$Y({\bf x};k)$ with $k = k_t$ at $\kappa_{\rm val} = 0.1340$ is presented in Fig.~\ref{fig:Y_40}.
The values of $Y({\bf x};k)$ at some positions in $x > 10$ are found to be close to zero,
satisfying $|Y({\bf x};k)| < 10^{-6}$.
Similar results of $Y({\bf x};k)$ are obtained in other $\kappa_{\rm val}$.
From the estimation we choose a reference point as ${\bf x}_{\rm ref} = (12,7,2)$.

The effective range expansion defines the scattering length $a_0$ and the effective range $r_{\rm eff}$,
\begin{equation}
 \frac{k}{\tan \delta(k)} = \frac{1}{a_0} + r_{\rm eff} k^2 + O(k^4).
 \label{eq:ERE}
\end{equation}
We estimate $a_0$ by
\begin{equation}
 a_0 = \frac{\tan \delta(k)}{k}.
 \label{eq:a_0}
\end{equation}
The tiny value of $k^2$ presented in Table~\ref{tab:k2_24x96} justifies the estimation.
Similarly, $r_{\rm eff}$ can be evaluated by
\begin{eqnarray}
 r_{\rm eff}
 &=& - \frac{2 k^2 H'/H_L(k;k) + \sin^2 \delta(k)}
            {2 k \sin \delta(k) \cos \delta(k)},
 \label{eq:r_eff}
 \\
 H'
 &=& \left.\frac{\partial H_L(p;k)}{\partial p^2}\right|_{p^2 = k^2} ,
\end{eqnarray}
where we assume
\begin{enumerate}
 \item The phase of $H(p;k)$ is $e^{i \delta(k)}$ at $p^2 \approx k^2$
 \item $\displaystyle
        \left.\frac{ \partial H(p;k) e^{-i \delta(k)} }{ \partial p^2 } \right|_{p^2 = k^2} =
        \left.\frac{ \partial H(p;p) e^{-i \delta(p)} }{ \partial p^2 } \right|_{p^2 = k^2}$ .
\end{enumerate}

\subsubsection{Scattering length}

\begin{figure}[t]
 \centering
 \includegraphics[scale=0.60]{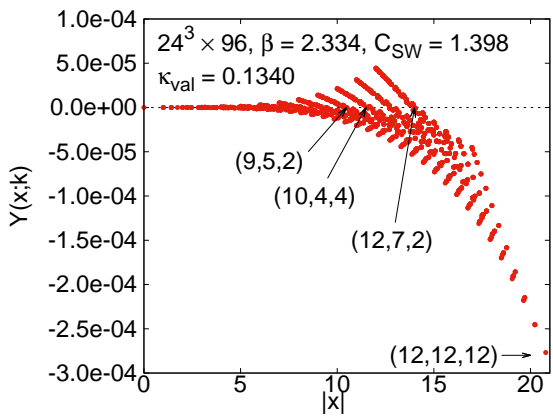}
 \caption{
  \label{fig:Y_40}
  $Y({\bf x};k)$ in Eq.~(\ref{eq:def_Y}) at $\kappa_{\rm val}=0.1340$ as a function of $|{\bf x}|$.
  Arrow expresses three components of each ${\bf x}$.
 }
\end{figure}

\begin{figure*}[t]
 \centering
 \includegraphics[scale=0.60]{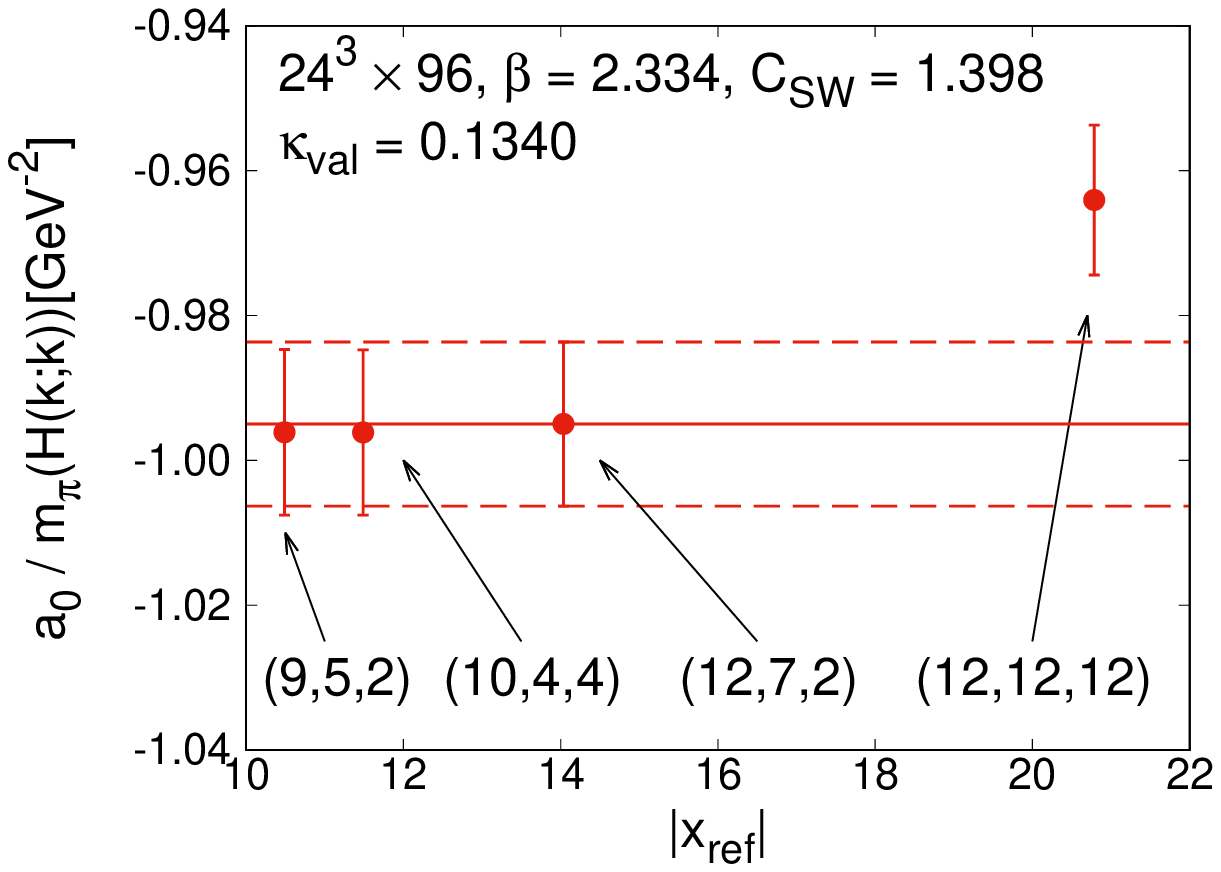}
 \includegraphics[scale=0.60]{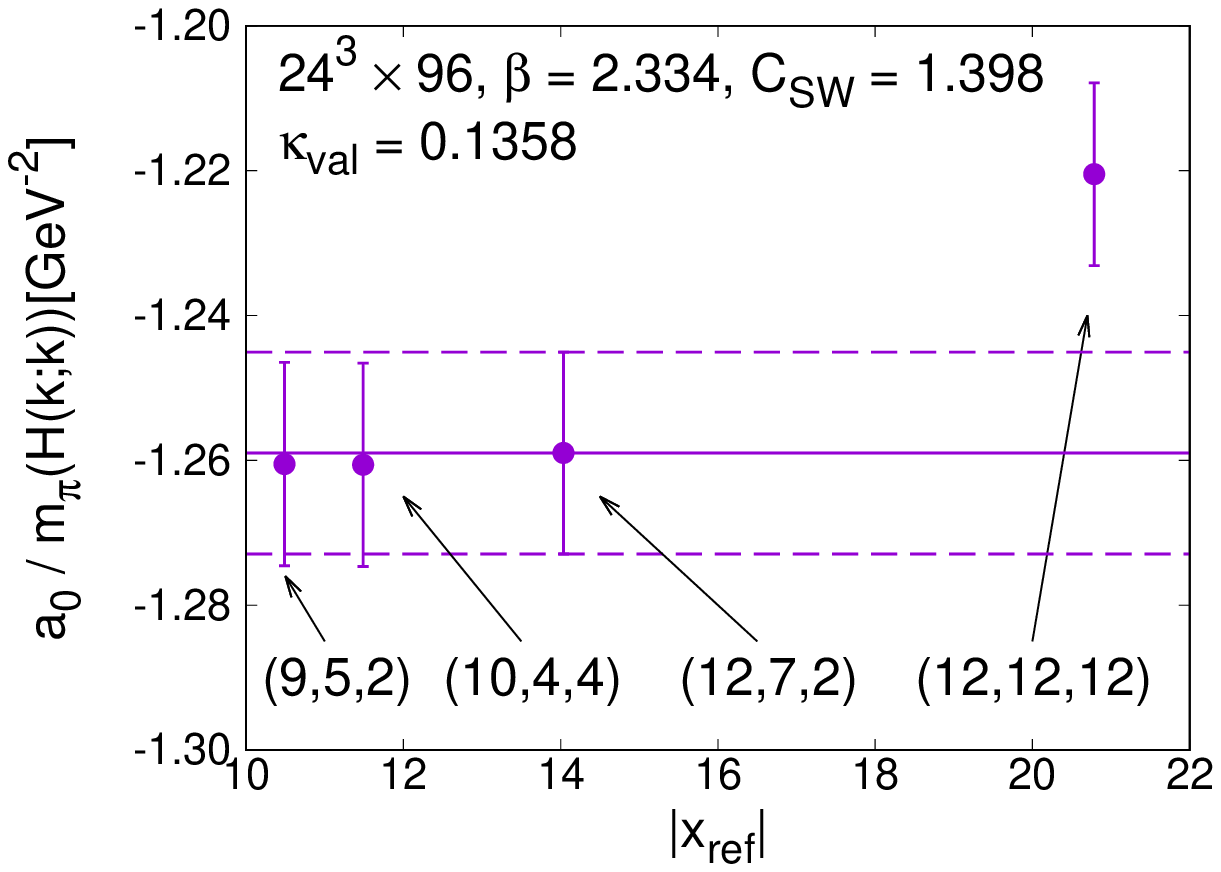}
 \includegraphics[scale=0.60]{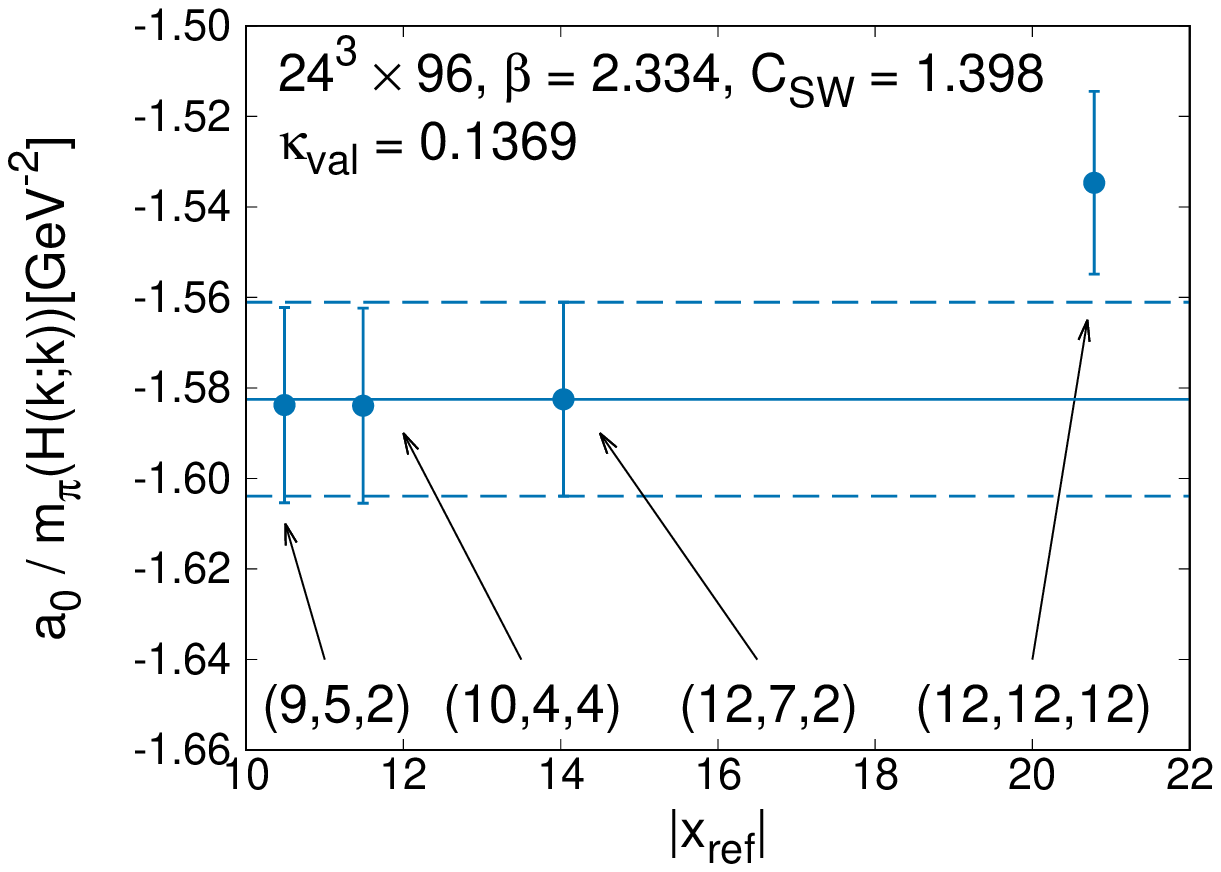}
 \caption{
  \label{fig:x_ref-dep}
  Reference point ${\bf x}_{\rm ref}$ dependence of $a_0 / m_\pi$ obtained from $H(k;k)$.
  The solid line represents the result with ${\bf x}_{\rm ref} = (12,7,2)$ with $1 \sigma$ error band.
 }
\end{figure*}

We evaluate $a_0 / m_\pi$ through Eq.~(\ref{eq:a_0}) using $\tan \delta(k)$ obtained from Eq.~(\ref{eq:tan_delta}).
In the evaluation, not only $k = k_s$ but also $k = k_t$ is employed.
A smaller error of $a_0 / m_\pi$ is obtained by $k = k_s$.
The results are tabulated in Table~\ref{tab:a_0}.

Since $\tan \delta(k)$ in Eq.~(\ref{eq:tan_delta}) depends on the choice of the reference point $x_{\rm ref}$,
$x_{\rm ref}$ dependence of $a_0 / m_\pi$ is also investigated.
Figure~\ref{fig:x_ref-dep} exhibits $x_{\rm ref}$ dependence of $a_0 / m_\pi$ at each $\kappa_{\rm val}$.
The left two data are obtained with the reference positions, 
${\bf x}_{\rm ref} = (9,5,2)$ and $(10,4,4)$, which satisfy the same condition $|Y({\bf x};k)| < 10^{-6}$
as ${\bf x}_{\rm ref} = (12,7,2)$, expressed in Fig.~\ref{fig:Y_40}.
These data are consistent with each other.
Contrarily, the rightest point in Fig.~\ref{fig:x_ref-dep}, ${\bf x}_{\rm ref} = (12,12,12)$,
overestimates the values from the other reference positions beyond 1 $\sigma$ error band.
${\bf x}_{\rm ref} = (12,12,12)$ gives the largest value of $|Y({\bf x};k)|$ as presented in Fig.~\ref{fig:Y_40}.
The result clearly indicates a sizable $l=4$ contribution at ${\bf x}_{\rm ref} = (12,12,12)$.
Our analysis of $x_{\rm ref}$ dependence of $a_0 / m_\pi$ suggests
the $l=4$ contribution in $a_0 / m_\pi$ is suppressed well by our choice of ${\bf x}_{\rm ref} = (12,7,2)$.

\begin{figure}[t]
 \centering
 \includegraphics[scale=0.60]{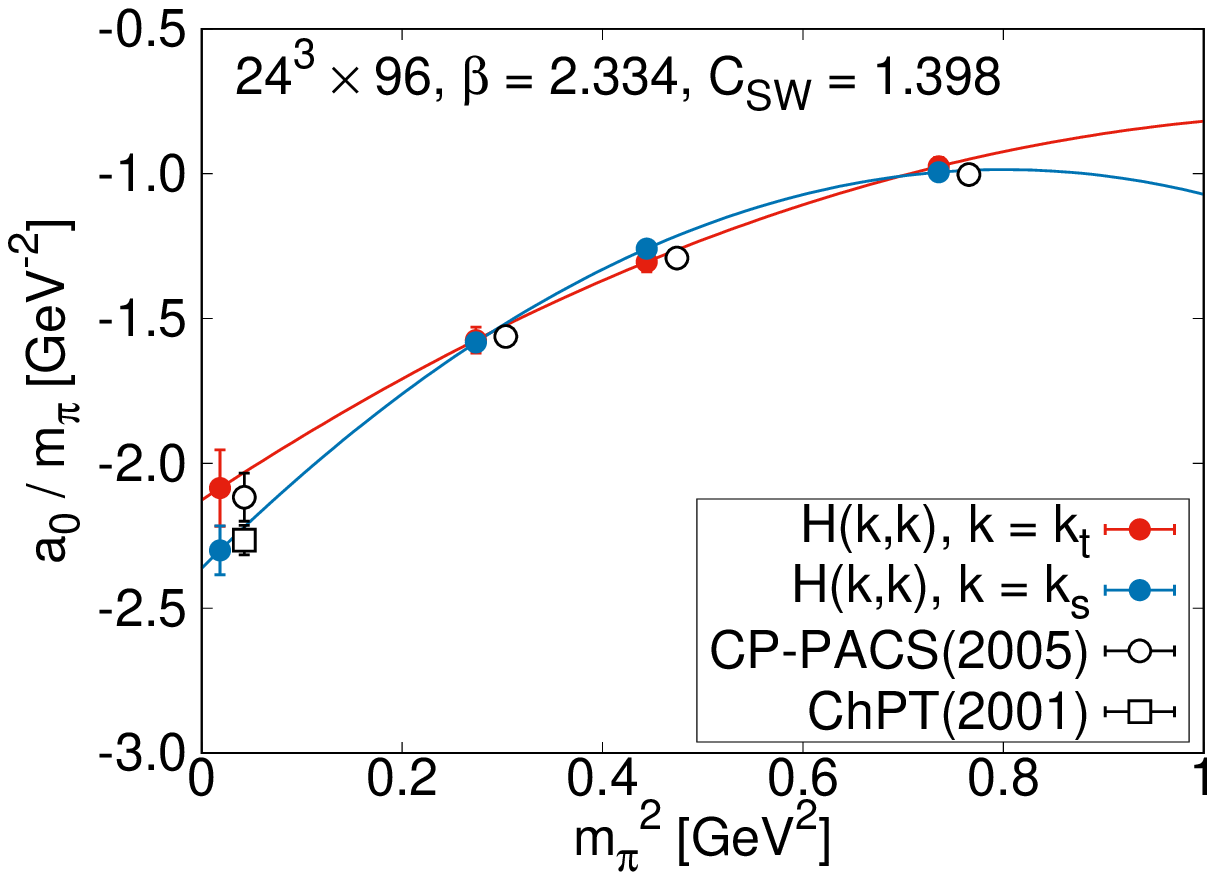}
 \caption{
  \label{fig:chiral_extrap_a_0}
  Quark mass dependence of the ratio of the scattering length over the pion mass $a_0 / m_\pi$.
  Open circles are lattice QCD results with the conventional finite volume method~\cite{Aoki:2005uf}.
  Open square is a phenomenological estimate of Chiral Perturbation Theory(ChPT)~\cite{Colangelo:2001df}.
  Open symbols are shifted to some extent for clarification of data.
 }
\end{figure}

$a_0 / m_\pi$ is extrapolated to the physical $m_\pi$
using a formula motivated by chiral perturbation theory~\cite{Gasser:1983yg},
\begin{eqnarray}
 \frac{a_0}{m_\pi}
 = A_{a_0} + B_{a_0} m_\pi^2 + C_{a_0} m_\pi^4,
\end{eqnarray}
where $A_{a_0},B_{a_0},C_{a_0}$ are fitting parameters.
Our results at the physical point are listed in Table~\ref{tab:a_0}.
Figure~\ref{fig:chiral_extrap_a_0} summaries chiral extrapolations of $a_0 / m_\pi$.
Our value at the physical point is consistent with the previous result by lattice QCD
using the conventional finite volume method based on $\phi({\bf x};k)$ outside the interaction range~\cite{Aoki:2005uf}
and the phenomenological estimate~\cite{Colangelo:2001df}.
The agreement ensures our approach with $\phi({\bf x};k)$ inside the interaction range.

\begin{table}[t]
\begin{center}
\begin{tabular}{ccc}
 \hline \hline
 $\kappa_{\rm val}$ &  $a_0 / m_\pi(k_t)$[GeV$^{-2}$]  &  $a_0 / m_\pi(k_s)$[GeV$^{-2}$]
 \\ \hline
 0.1340             & $-0.975(31)$ & $-0.995(11)$
 \\
 0.1358             & $-1.305(35)$ & $-1.259(14)$
 \\
 0.1369             & $-1.575(45)$ & $-1.582(21)$
 \\\hline
 Physical           & $-2.09(13)$  & $-2.30(8)$
 \\ \hline \hline 
\end{tabular}
 \caption{
  \label{tab:a_0}
  Scattering length $a_0$ over the pion mass $m_\pi$ obtained with $k_t$ and $k_s$ on $24^3 \times 96$.
 }
\end{center}
\end{table}

\subsubsection{Effective range}

\begin{figure}[t]
 \centering
 \includegraphics[scale=0.60]{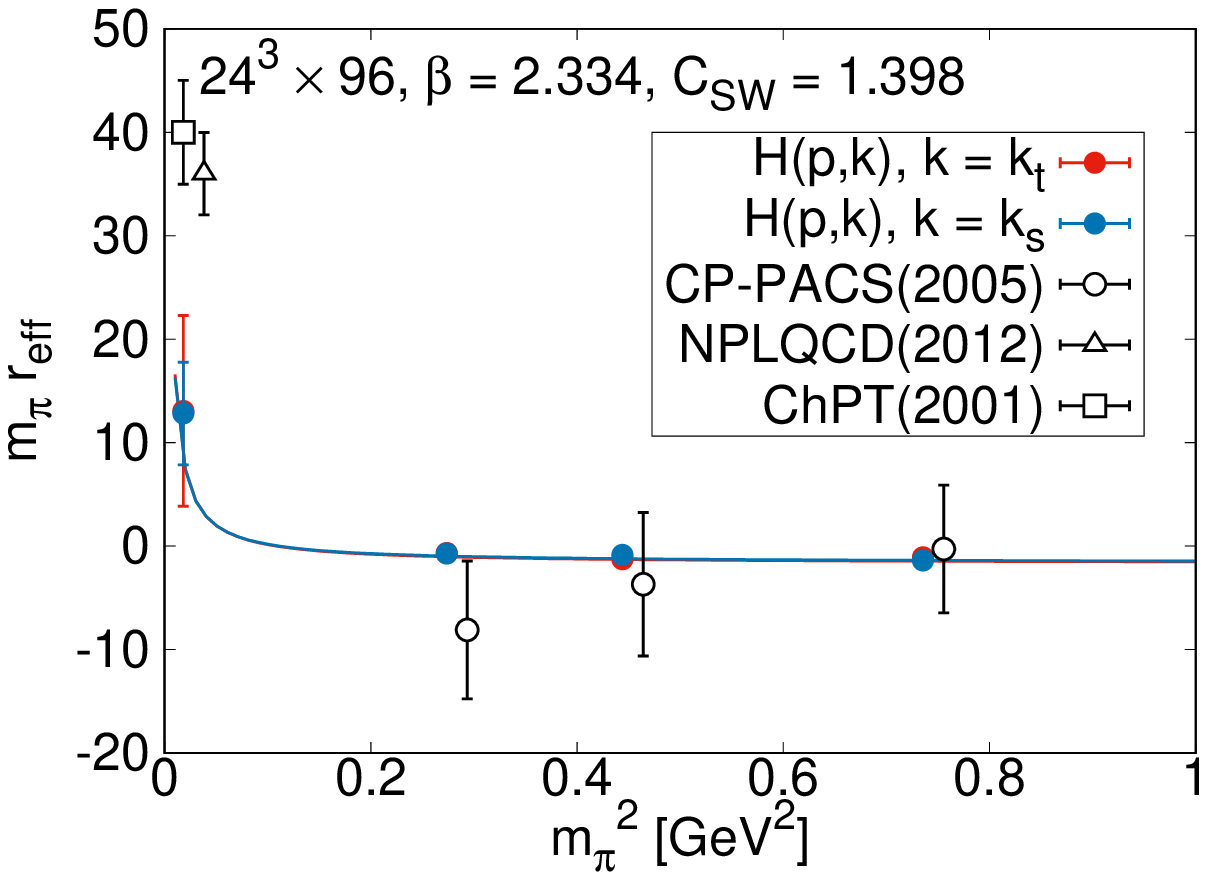}
 \caption{
  \label{fig:chiral_extrap_r_eff}
  Quark mass dependence of the effective range $r_{\rm eff}$.
  Open circles are $N_f=0$ lattice QCD results with the conventional finite volume method
  calculated using data in Ref.~\cite{Aoki:2005uf}.
  Open triangle is $N_f=2+1$ lattice QCD result with the finite volume method~\cite{Beane:2011sc}.
  Open square is a phenomenological estimate of Chiral Perturbation Theory(ChPT)~\cite{Colangelo:2001df}.
  Open symbols are shifted to some extent for clarification of data.
 }
\end{figure}

The effective range $r_{\rm eff}$ is evaluated by the slope of $H_L(p;k)$ with respective to $p^2$ and $\delta(k)$ 
shown in Eq.~(\ref{eq:r_eff}).
Our results with $k = k_s$ and $k = k_t$ are compiled in Table~\ref{tab:r_eff}.
The result using $k_t$ has a larger error than that using $k_s$, as in the case of $a_0$.
Figure~\ref{fig:chiral_extrap_r_eff} plots $m_\pi$ dependence of $m_\pi r_{\rm eff}$.
Our result of $m_\pi r_{\rm eff}$ agrees with the value by the finite volume method,
calculated using data in Ref.~\cite{Aoki:2005uf}.
Our data are more accurate than those of the finite volume method
due to the explicit $p^2$ dependence of $H(p;k)$.

$r_{\rm eff}$ is extrapolated to the physical point
using a formula based on chiral perturbation theory~\cite{Bijnens:1997vq}.
\begin{eqnarray}
 m_\pi r_{\rm eff}
 = \frac{ A_{r_{\rm eff}} }{ m_\pi^2 } + B_{r_{\rm eff}},
\end{eqnarray}
where $A_{r_{\rm eff}},B_{r_{\rm eff}}$ are fitting parameters.
Our result at the physical point, whose value is summarized in Table~\ref{tab:r_eff}, 
underestimates the phenomenological value~\cite{Colangelo:2001df}.
The reason seems to be the chiral extrapolation of $m_\pi r_{\rm eff}$.
It rapidly grows toward the physical point.
We also need to validate the two assumptions in Eq.~(\ref{eq:r_eff}),
though consistency between our result and that of the conventional finite volume method is confirmed
at each simulation point.
Another possibility is the quenching effect.
In fact, $N_f=2+1$ lattice QCD using $m_\pi = 390$~MeV
successfully reproduces the phenomenological estimate~\cite{Beane:2011sc}.
More realistic $N_f=2+1$ data around the physical point are required to draw a definite conclusion.

\begin{table}[t]
\begin{center}
\begin{tabular}{ccc}
 \hline \hline
 $\kappa_{\rm val}$  &  $r_{\rm eff}(k_t)$[GeV$^{-1}$]  &  $r_{\rm eff}(k_s)$[GeV$^{-1}$]
 \\ \hline
 0.1340              &  $-1.26(63)$  &  $-1.63(22)$
 \\
 0.1358              &  $-1.90(49)$  &  $-1.28(25)$
 \\
 0.1369              &  $-1.26(64)$  &  $-1.36(28)$
 \\ \hline
 Physical            &  $10.8(7.6)$  &  $10.6(4.1)$
 \\ \hline \hline 
\end{tabular}
 \caption{
  \label{tab:r_eff}
  Effective ranges $r_{\rm eff}$ using $k_t$ and $k_s$ on $24^3 \times 96$.
 }
\end{center}
\end{table}

\section{Summary}
\label{sec:summary}

We have successfully calculated the on-shell and half off-shell scattering amplitudes
from the BS wave function inside the interaction range,
as reported in our previous paper~\cite{Namekawa:2017sxs}.
Our approach utilizes the BS wave function inside the interaction range,
while the conventional finite volume method is based on the BS wave function outside the interaction range.
The on-shell scattering amplitude gives direct scattering information through the phase shift.
The half off-shell amplitude is not an observable in experiments, on the other hand,
but it could be an important input of theoretical effective theories and models to constrain their parameters.
Furthermore, the half off-shell amplitude gives the effective range under two assumptions.

In this article, we extended our study to investigate quark mass dependence of $I=2$ S-wave two-pion on-shell
and half off-shell scattering amplitudes at the center of mass in the quenched QCD.
Our simulation was performed at the lattice spacing of $a^{-1} = 1.207$~GeV
using pion masses of $m_\pi = 0.52-0.86$~GeV on $24^3 \times 96$ lattice.
We first checked the interaction range is within half of our spatial lattice,
which satisfies a sufficient condition of our method as well as the finite volume method.
It allows us to evaluate on-shell and half off-shell scattering amplitudes.
We obtained clean signals of them.
We also discussed the source and sink operator independence of our scattering amplitudes.

We then extracted the scattering length from the on-shell scattering amplitude.
Our results at each $m_\pi$ and the physical point agree with those obtained by the finite volume method.
It proves our approach is an alternative to the conventional finite volume method.

We also extracted the effective range from the slope of the half off-shell amplitude
at the on-shell momentum under two assumptions to be validated.
Our result agrees with that by the conventional finite volume method at each pion mass.
On the physical point, however, our result extrapolated from data with pion masses of $0.52-0.86$~GeV
underestimates the recent lattice QCD and the phenomenological values.
More realistic data near the physical point in $N_f=2+1$ lattice QCD are required
to identify the reason of the underestimation.

\begin{acknowledgments}

We thank J.~Carbonell and V.~A.~Karmnov for pointing out the momentum space formulation,
and members of PACS collaboration for useful discussion.
We also appreciate INT at University of Washington for the kind hospitality and partial financial support
during an early stage of this work.
Our simulation was performed on COMA
under Interdisciplinary Computational Science Program of Center for Computational Sciences, University of Tsukuba.
This work is based on Bridge++ code (http://bridge.kek.jp/Lattice-code/)~\cite{Ueda:2014rya}.
This work is supported in part by JSPS KAKENHI Grant Numbers 16H06002 and 18K03638.

\end{acknowledgments}

\appendix

\section{Operator dependence of scattering amplitude}
\label{app:operator-dep}

We discuss source and sink operator dependence of the scattering amplitude.

\subsection{Source operator dependence}

\begin{table}[tb]
\begin{center}
\begin{tabular}{ccccc}\hline\hline
 Lattice size     &  $\kappa_{\rm val}$    &  $N_{\rm src}$  &  Source type  &  $N_{\rm config}$
 \\ \hline
 $24^3 \times 64$ &  0.1340                &  32             &  Z2           &  400
 \\
                  &                        &  16             &  Wall         &  200
 \\ \hline \hline
\end{tabular}
 \caption{
  \label{tab:simulation_24x64}
  Simulation parameters on $24^3 \times 64$ lattice.
 }
\end{center}
\end{table}

\begin{figure}[t]
 \centering
 \includegraphics[scale=0.60]{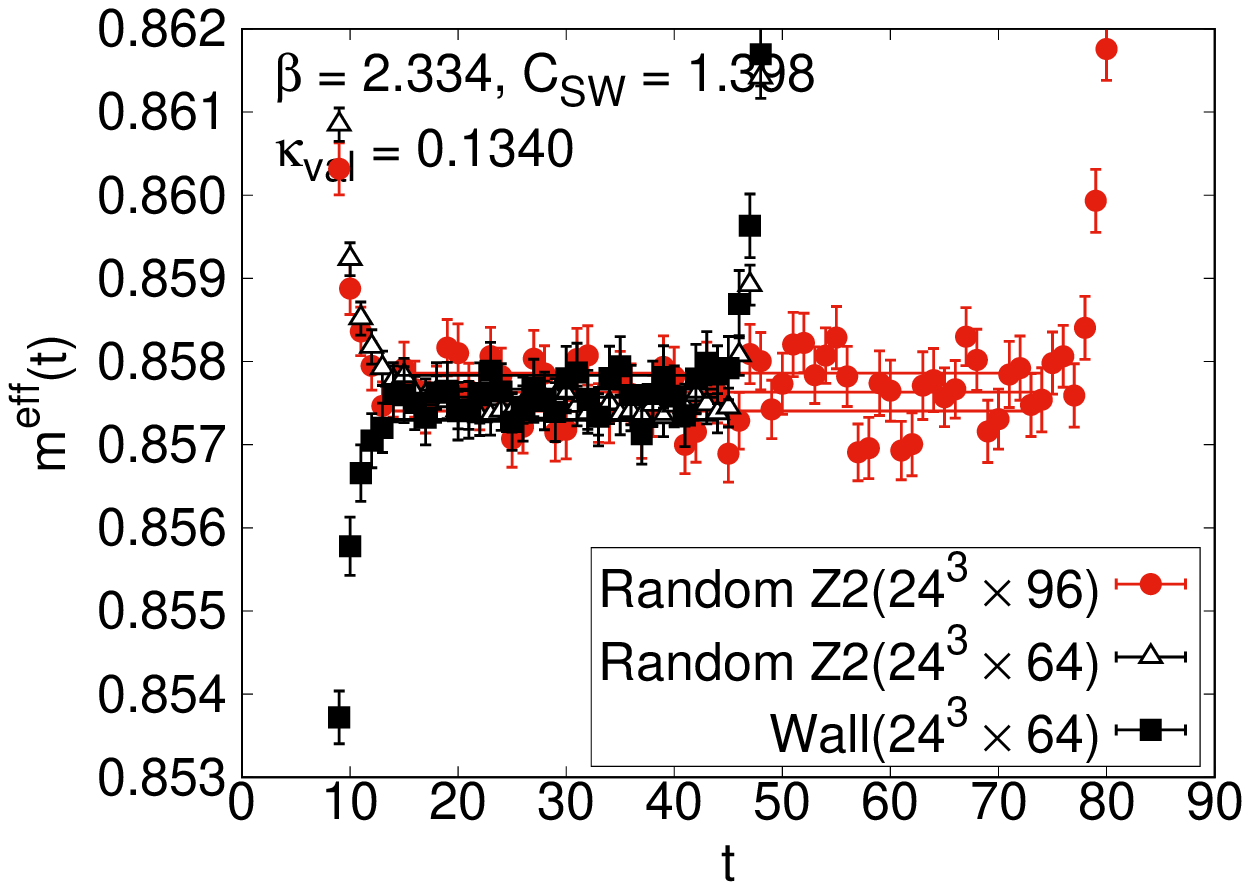}
 \includegraphics[scale=0.60]{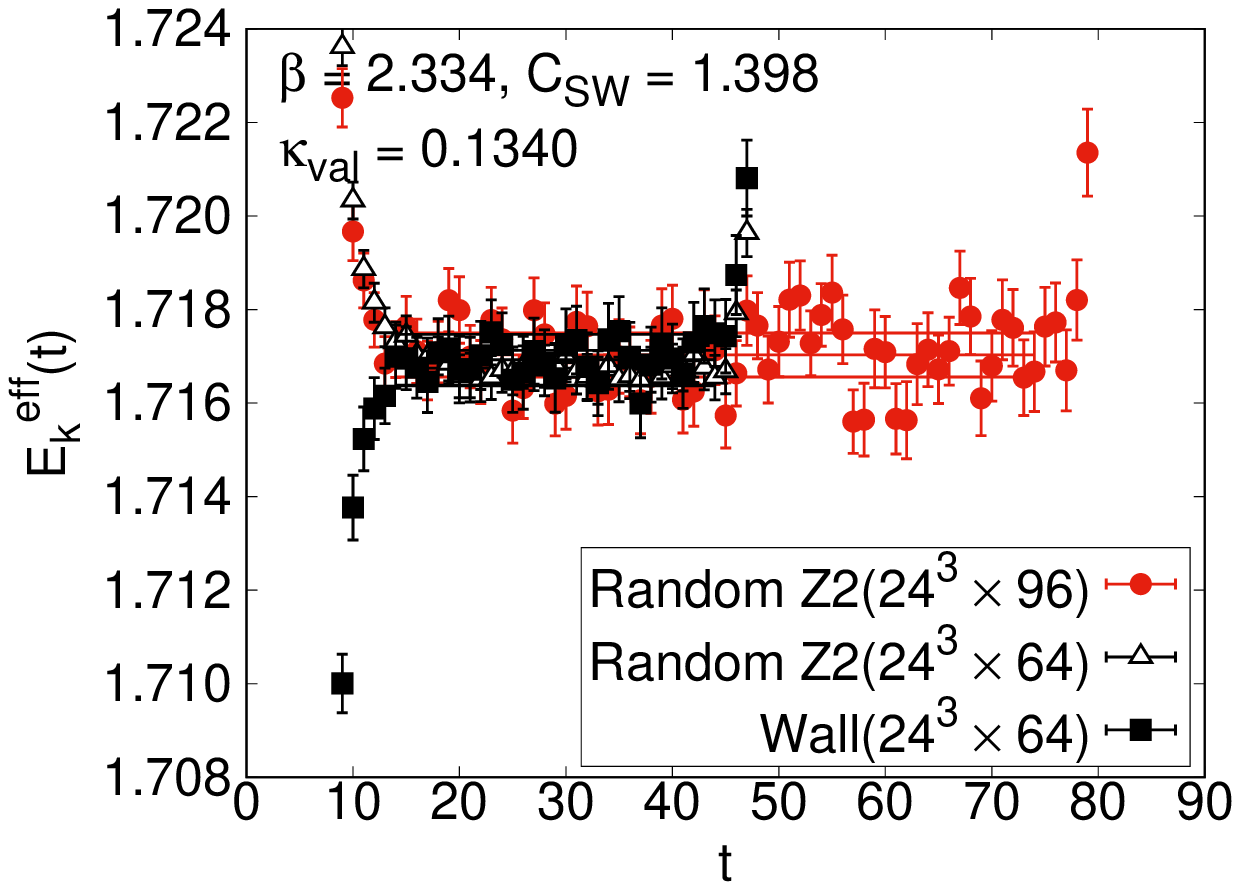}
 \caption{
  \label{fig:eff_W}
  Effective masses and two-pion energies with random Z2 and wall sources on $24^3 \times 64$ lattice.
  The data with random Z2 source on $24^3 \times 96$ are also plotted.
 }
\end{figure}

\begin{figure}[t]
 \centering
 \includegraphics[scale=0.60]{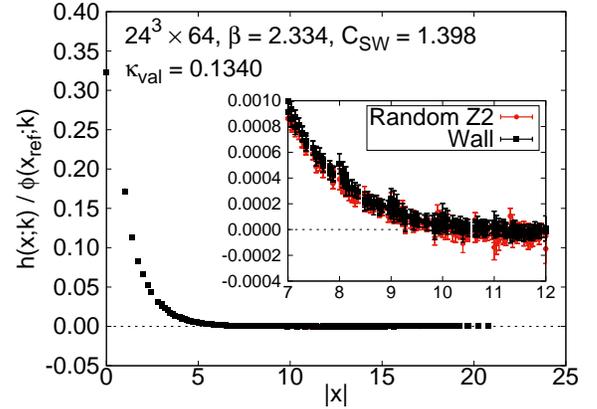}
 \caption{
  \label{fig:h_W_kt}
  Comparison of $h({\bf x};k)$ using $k = k_t$ with random Z2 and wall sources on $24^3 \times 64$ lattice.
 }
\end{figure}

\begin{table}[t]
\begin{center}
\begin{tabular}{ccc}
 \hline \hline
 source           & Z2                         & Wall
 \\ \hline
 $m_\pi$[GeV]     & 0.85748(19)                & 0.85757(26)
 \\
 $E_k$[GeV]       & 1.71675(38)                & 1.71693(54)
 \\
 $k_t^2$[GeV$^2$] & 1.533(34) $\times 10^{-3}$ & 1.535(56) $\times 10^{-3}$
 \\
 $k_s^2$[GeV$^2$] & 1.569(27) $\times 10^{-3}$ & 1.523(27) $\times 10^{-3}$
 \\
 $a_0/m_\pi(k_s)$[GeV$^{-2}$]
                  & $-$0.963(10) & $-$0.979(10)
 \\
 $r_{\rm eff}(k_s)$[GeV$^{-2}$]
                  & $-$1.85(10)  & $-$1.73(09) 
 \\ \hline \hline 
\end{tabular}
 \caption{
  \label{tab:k2_24x64}
  $m_\pi$, $E_k$, $k_t^2$, and $k_s^2$ on $24^3 \times 64$ using random Z2 and wall sources at $k_{\rm val} = 0.1340$.
 }
\end{center}
\end{table}

The source operator dependence is simply explained by an overall factor of $\phi({\bf x};k)$.
When the ground state dominates $C_{\pi\pi}(t)$ in large $t$ region, the source operator dependence cancels in ratios,
$\phi({\bf x};k) / \phi({\bf x}_{\rm ref};k)$ and $h({\bf x};k) / \phi({\bf x}_{\rm ref};k)$.

In order to check the source operator independence of $h({\bf x};k)/\phi({\bf x}_{\rm ref};k)$,
we compare results using the Z2 random and wall sources on the $24^3 \times 64$ ensemble with $\kappa_{\rm val} = 0.1340$,
which was used partly in our previous paper~\cite{Namekawa:2017sxs}.
The same simulation set up is adopted as that in Sec.~\ref{sec:set_up}.
We use 400(200) configurations with 32(16) measurements per configuration using the random Z2(wall) source. 
The simulation parameters are listed in Table~\ref{tab:simulation_24x64}.

The wall source operator $\Omega_{\rm wall}(t)$ in $C_{\pi\pi}({\bf x},t)$ of Eq.~(\ref{eq:4point_func}) is defined by
\begin{eqnarray}
 \Omega_{\rm wall}(t)
 = \pi^+_{\rm wall}(t_{\rm src}) \pi^+_{\rm wall}(t_{\rm src} + 1),
\end{eqnarray}
where
\begin{equation}
 \pi^+_{\rm wall}(t_{\rm src})
 = \left[\sum_{{\bf x}_1}\overline{d}({\bf x}_1,t)\right]
   \gamma_5
   \left[\sum_{{\bf x}_2} u({\bf x}_2,t)\right] .
\end{equation}
The pion operators in $\Omega_{\rm wall}(t)$ are placed at $t_{\rm src}$ and $t_{\rm src} + 1$
to prevent from Fierz rearrangement~\cite{Kuramashi:1993ka,Fukugita:1994ve}.

Figure~\ref{fig:eff_W} exhibits the effective mass $m_\pi^{\rm eff}$ and energy $E_k^{\rm eff}$,
defined in Eqs.~(\ref{eq:m_eff}) and (\ref{eq:E_eff}) respectively.
Our results from the two sources are consistent with each other,
as well as that on $24^3 \times 96$ lattice.
We fit data in the range of $[t_{\rm min},t_{\rm max}] = [14,44]$ to extract $m_\pi$ and $E_k$.
The results are summarized in Table~\ref{tab:k2_24x64} together with those for $k^2_t$ and $k^2_s$.

In contrast to the case on $24^3 \times 96$ lattice in Fig.~\ref{fig:wavefunc_eff_24x96},
the data of $\phi({\bf x};k)$ on $24^3 \times 64$ lattice available for analysis of the BS wave function
are limited to those at a time slice of $t = 44$.
We use $\phi({\bf x};k)$ at $t = 44$ for our analysis on $24^3 \times 64$ lattice.

Figure~\ref{fig:h_W_kt} presents the two source results of $h({\bf x};k) / \phi({\bf x}_{\rm ref};k)$.
They agree with each other, and with the result obtained on $24^3 \times 96$ lattice in Fig.~\ref{fig:reduced_phi_L}.
Consistency of the two results proves the source independence of $h({\bf x};k_s) / \phi({\bf x}_{\rm ref};k)$.
We also confirmed the source operator independence of the physical quantities $a_0 / m_\pi$ and $r_{\rm eff}$ 
extracted from $H_L(p;k)$.
The results are in Table~\ref{tab:k2_24x64}.
Our analysis establishes the source operator dependence is well under control.

\subsection{Sink operator dependence}

The sink smearing of the pion operator in $C_{\pi\pi}(t)$ produces an extra overall factor.
It can not be removed by the same ratio as that in the source operator case.
Using a smearing function of the pion $f(x)$,
the sink smeared BS wave function $\bar{\phi}({\bf x};k)$ is obtained by
\begin{eqnarray}
 \bar{\phi}({\bf x};k) = \int d^3 y \, f(|{\bf x} - {\bf y}|) \phi({\bf y};k).
 \label{eq:phibar}
\end{eqnarray}
where only one of the pion operators is smeared.
$\bar{\phi}({\bf x};k)/\bar{\phi}({\bf x}_{\rm ref};k)$ is not the same as $\phi({\bf x};k)/\phi({\bf x}_{\rm ref};k)$
in contrast to the source operator case~\cite{Yamazaki:2017gjl}.

Nevertheless, the extra sink smearing factor can be analytically removed~\cite{Kawai:2017goq}.
The sink smeared scattering amplitude $\bar{H}_L(p;k)$ defined by $\bar{\phi}({\bf x};k)$ relates to $H_L(p;k)$ such that
\begin{eqnarray}
 \bar{H}_L(p;k)
 &=& - \int d^3 x \, (\Delta + k^2) \bar{\phi}({\bf x};k) e^{-i {\bf p} \cdot {\bf x}}
 \nonumber \\
 &=& C_f(p) H_L(p;k),
 \label{eq:hbar}
\end{eqnarray}
where
\begin{equation}
 C_f(p) = \int d^3 x \, f(x) e^{-i {\bf p} \cdot {\bf x}} .
 \label{eq:cf}
\end{equation}
The local operator corresponds to $f(x) = \delta^{(3)}(x)$ with $C_f(p) = 1$ in all $p$.
Once $f(x)$ is given, $C_f(p)$ can be analytically calculated and removed from $\bar{H}_L(p;k)$.

We numerically check the sink smearing independence of our result.
$\bar{\phi}({\bf x};k)$ and $\bar{H}_L(p;k)$ are calculated
by replacing the integration and $e^{-i {\bf p} \cdot {\bf x}}$ in Eqs.~(\ref{eq:phibar}), (\ref{eq:hbar}), 
and (\ref{eq:cf}) to the summation and $j_0(p x)$.
We employ an exponential sink smearing function $f(x) = e^{-A x}$ with a constant $A$.
We confirm the values of $\bar{H}_L(p;k) / C_f(p)$ with different $f(x)$ are consistent,
unless $f(x)$ is too broad comparing to $L/2 - R$.
A broad sink smearing function is found to lift up $h({\bf x};k)$ in $|{\bf x}| > R$,
which violates the sufficient condition Eq.~(\ref{eq:condition_h}).

\section{Formulation in the momentum space}
\label{sec:formulation_momentum_space}

Formulation of $H(p;k)$ using the BS wave function in momentum space is summarized.
We first explain the formulation in the infinite volume and continuum theory, and then present its lattice version.

The main difference from those in explained in Sec.~\ref{sec:formulation_lattice} is appearance of the surface term.
Only the surface term contributes to the on-shell scattering amplitude.

\subsection{Infinite volume}

The scattering amplitude with the infinite volume $H(p;k)$ in the continuum theory is given by
\begin{eqnarray}
 H(p;k) = - \int_{-\infty}^{\infty} d^3 x \, h_\infty({\bf x};k) e^{-i {\bf p} \cdot {\bf x}} .
 \label{eq:appendix_def_H_pk}
\end{eqnarray}
The reduced wave function $h_\infty({\bf x};k)$ is defined using the BS wave function
in the infinite volume $\phi_{\infty}({\bf x};k)$,
\begin{eqnarray}
 h_\infty({\bf x};k) = (\Delta + k^2) \phi_{\infty}({\bf x};k),
\label{eq:h_infty}
\end{eqnarray}
where $\Delta$ is the laplacian.
Substituting Eq.~(\ref{eq:h_infty}) to Eq.~(\ref{eq:appendix_def_H_pk}) with the partial integration yields
\begin{equation}
 H(p;k) = (p^2-k^2) \widetilde{\phi}_\infty({\bf p};k) .
 \label{eq:Hpk_tilde_phi}
\end{equation}
$\widetilde{\phi}_\infty({\bf p};k)$ is the BS wave function in the momentum space,
\begin{equation}
 \widetilde{\phi}_\infty({\bf p};k)
 = \int_{-\infty}^{\infty} d^3 x \, \phi_\infty({\bf x};k) e^{-i {\bf p} \cdot {\bf x}} .
 \label{eq:tilde_phi_infty}
\end{equation}
Equation~(\ref{eq:Hpk_tilde_phi}) could be regarded as the LSZ reduction formula in the relative coordinate.
It is constructed by the Fourier transformation of the BS wave function with a momentum factor.
It corresponds to the LSZ reduction formula of Eq.~(\ref{eq:LSZ}).
At on-shell, both formulae give the same on-shell scattering amplitude.
In the zero momentum limit, Eq.~(\ref{eq:Hpk_tilde_phi}) leads to the scattering length $a_0$,
as in Ref.~\cite{Carbonell:2016ekx}.
Equation~(\ref{eq:Hpk_tilde_phi}) is not suitable, however, for the lattice calculation on a finite volume.
$\phi_{\infty}({\bf x};k)$ at all ${\bf x}$ in the infinite volume is demanded.

In the following, we consider a formulation with the finite integration range~\cite{Namekawa:2017sxs}.
If $h_\infty({\bf x};k) = 0$ outside the interaction range $R$,
then the integration range of $H(p;k)$ can be changed from $\infty$ to $R$,
\begin{eqnarray}
 H(p;k)
 &=& - \int_{-\infty}^{\infty} d^3 x \, h_\infty({\bf x};k) e^{-i {\bf p} \cdot {\bf x}},
 \\
 &=& - \int_{-R}^{R} d^3 x \, h_\infty({\bf x};k) e^{-i {\bf p} \cdot {\bf x}}.
\end{eqnarray}
The partial integration gives
\begin{eqnarray}
 H(p;k)
 &=& (p^2 - k^2) \int_{-R}^{R} d^3 x \, \phi_{\infty}({\bf x};k) e^{-i {\bf p} \cdot {\bf x}}
 \nonumber \\
 && - \sum_{i=1}^{3} \int_{-R}^{R} d^2 x
                     \left[ \partial_i \phi_{\infty}({\bf x};k)
                                   + i p_i    \phi_{\infty}({\bf x};k)
                     \right]_{x_i = R},
 \nonumber \\
\end{eqnarray}
where the second term is the surface term.
At on-shell $p = k$, the first term vanishes.
$H(p;k)$ is expressed by the surface term only.

In the spherical coordinate, $H(p;k)$ can be simplified.
\begin{eqnarray}
 H(p;k)
 &=& 4 \pi (p^2 - k^2) \int_{0}^{R} dx x^2 \, \phi_{\infty}({\bf x};k) j_0(p x)
 \nonumber \\
 && - \frac{4 \pi}{p}
      \left\{
        R \sin(p R) \frac{\partial \phi_{\infty}({\bf x};k)}{\partial x} \middle|_{x = R}
      \right.
 \nonumber \\
 && - \left.
        \left(
          p R \cos(p R) - \sin(p R)
        \right) \phi_{\infty}(R;k)
      \right\}.
\end{eqnarray}
At on-shell, the expression of $H(k;k)$ in Eq.~(\ref{eq:H_kk_infinite_volume}) is reproduced
by substituting the following $l=0$ $\phi_{\infty}({\bf x};k)$ in $x > R$ to the surface term,
\begin{equation}
 \phi_{\infty}({\bf x};k) = e^{i \delta(k)} \frac{ \sin( k x + \delta(k) )}{k x} .
 \label{eq:phi_infty_delta}
\end{equation}

\subsection{Finite volume}

The lattice version of Eq.~(\ref{eq:appendix_def_H_pk}) is
\begin{eqnarray}
 H_L(p;k)
 = - \sum_{i = 1}^{3}
     \sum_{x_i = -L_{\rm min}}^{L_{\rm max}}
       C_k h({\bf x};k) e^{-i {\bf p} \cdot {\bf x}},
\end{eqnarray}
where $C_k$ is an overall constant in Eq.~(\ref{eq:H_L_pk_lattice}).
$h({\bf x};k)$ is the reduced wave function on the lattice.
A choice of $L_{\rm max} = L/2$ and $L_{\rm min} = L/2 - 1$ corresponds to
the summation over the entire spatial volume with its extent $L$.
$L_{\rm max}$ and $L_{\rm min}$ can be decreased,
as long as $L_{\rm max}, L_{\rm min} > R$ is satisfied.

The partial integration on the lattice leads to
\begin{eqnarray}
 H_L(p;k)
 &=& (\tilde{p}^2 - k^2)
     \sum_{i = 1}^{3} \sum_{x_i = -L_{\rm min}}^{L_{\rm max}}
     C_k \phi({\bf x};k) e^{-i {\bf p} \cdot {\bf x}}
 \nonumber \\
 && + \mbox{surf}(p;k),
 \label{eq:HL_mom_exp}
\end{eqnarray}
where
\begin{eqnarray}
 \tilde{p}_i
 = \frac{2}{a} \sin \frac{a p_i}{2} .
\end{eqnarray}
The surface term on the lattice $\mbox{surf}(p;k)$ is given by
\begin{eqnarray}
 \mbox{surf}(p;k)
 &=& - C_k
     \sum_{i=1}^{3} 
     \sum_{\substack{x_{j,k} = -L_{\rm min} \\ j,k \ne i}}^{L_{\rm max}}
 \nonumber \\
 &&  \left( e^{-i {\bf p} \cdot {\bf X}(L_{\rm max})}\phi({\bf X}(L_{\rm max}+1);k) 
     \right.
 \nonumber \\
 && - e^{-i {\bf p} \cdot {\bf X}(L_{\rm max}+1)}\phi({\bf X}(L_{\rm max});k)
 \nonumber \\
 && - e^{-i {\bf p} \cdot {\bf X}(-L_{\rm min}-1)} \phi({\bf X}(-L_{\rm min});k) 
 \nonumber \\
 && \left.+ e^{-i {\bf p} \cdot {\bf X}(-L_{\rm min})} \phi({\bf X}(-L_{\rm min}-1);k)
    \right),
 \nonumber \\
\end{eqnarray}
where ${\bf X}(a) = {\bf x}$ except for $X_i(a) = a$.
The surface term is not zero, in general.
If $L_{\rm max} = L/2$ and $L_{\rm min} = L/2 - 1$ are chosen and
$p_i L_{\rm max} = n_i \pi, n_i \in {\bf Z}$ is satisfied
under the periodic boundary condition, the surface term becomes zero.

In the case of the S-wave scattering on the lattice, $H_L(p;k)$ becomes
\begin{eqnarray}
 H_L(p;k)
 &=& -k^2 \sum_{i = 1}^{3} \sum_{x_i = -L_{\rm min}}^{L_{\rm max}} C_k \phi({\bf x};k) j_0(p x)
 \nonumber \\
 &&  -    \sum_{i = 1}^{3} \sum_{x_i = -L_{\rm min}}^{L_{\rm max}} C_k \phi({\bf x};k) \Delta j_0(p x)
 \nonumber \\
 &&  + \mbox{surf}(p;k),
 \label{eq:HL_mom_j0}
\end{eqnarray}
where $\Delta$ is the symmetric lattice laplacian defined in Eq.~(\ref{eq:lat_laplacian}).
The surface term in this case is given by
\begin{eqnarray}
 \mbox{surf}(p;k)
 &=& - C_k
     \sum_{i=1}^{3}
     \sum_{\substack{x_{j,k} = -L_{\rm min} \\ j,k \ne i}}^{L_{\rm max}}
 \nonumber \\
 &&  \left( j_0(p X(L_{\rm max}))\phi({\bf X}(L_{\rm max} + 1);k) 
     \right.
 \nonumber \\
 && - j_0(p X(L_{\rm max}+1))\phi({\bf X}(L_{\rm max});k)
 \nonumber \\
 && - j_0(pX(-L_{\rm min}-1)) \phi({\bf X}(-L_{\rm min});k) 
 \nonumber \\
 && \left.+ j_0(p X(-L_{\rm min})) \phi({\bf X}(-L_{\rm min} - 1);k)
    \right),
 \nonumber \\
\end{eqnarray}
where $X(a) = |{\bf X}(a)|$.
If $L_{\rm max} = L/2$ and $L_{\rm min} = L/2 - 1$ are chosen,
the periodicity and isotropy of $\phi({\bf x};k)$ leads to a simpler form of $\mbox{surf}(p;k)$,
\begin{eqnarray}
 \mbox{surf}(p;k)
 &=& - 3 C_k
     \sum_{x_{1,2} = -L_{\rm min}}^{L_{\rm max}}
 \nonumber \\
 && [ j_0(pX^\prime(L_{\rm min}))-j_0(pX^\prime(L_{\rm max}+1)) ]
 \nonumber \\
 && \times \phi({\bf X}^\prime(L_{\rm max});k) ,
 \label{eq:surface_j0}
\end{eqnarray}
where $X_{1,2}^\prime(a) = x_{1,2}$ and $X_3^\prime(a) = a$.

\section{$t$ independence of $H_L(k;k) / (C_k \phi({\bf x}_{\rm ref};k))$}
\label{app:ratio}

The $t$ dependence of $H_L(k;k) / (C_k \phi({\bf x}_{\rm ref};k))$ in Fig.~\ref{fig:t_dependence_of_H_over_phi}
is discussed under several assumptions.

We define $H_L(t,k;k)$ to study the scattering amplitude $H_L(k;k)$ at each $t$,
\begin{eqnarray}
 H_L(t,k;k)
 &=& - \sum_{\bf x} (\Delta + k^2) C_{\pi\pi}({\bf x},t) j_0(k x) .
 \label{eq:HL_tkk}
\end{eqnarray}
We evaluate a ratio of $H_L(t,k;k) / C_{\pi\pi}({\bf x},t)$.
It can be split into the ground state and the excited state parts,
\begin{equation}
 \frac{H_L(t,k;k)}{C_{\pi\pi}({\bf x}_{\rm ref},t)}
 =
 \frac{H_L(k;k)}{C_k \phi({\bf x}_{\rm ref};k)}
 \frac{1 + \delta H_L(t,k;k)}{1 + \delta C_{\pi\pi}({\bf x}_{\rm ref},t)} ,
 \label{eq:ratio_HL_phi}
\end{equation}
where the excited state contributions $\delta H_L(t,k;k)$ and $\delta C_{\pi\pi}({\bf x}_{\rm ref},t)$
for the numerator and denominator, respectively.
Figure~\ref{fig:t_dependence_of_H_over_phi} illustrates
$H_L(t,k;k) / C_{\pi\pi}({\bf x}_{\rm ref},t)$ is almost flat against $t$.
On the other hand, Fig.~\ref{fig:each_t_H_phi} clearly reveals
non-negligible contributions of $\delta H_L(t,k;k)$ and $\delta C_{\pi\pi}({\bf x}_{\rm ref},t)$ in the small $t$ region.
It suggests a possibility of cancellation between the numerator and the denominator,
\begin{equation}
 \delta H_L(t,k;k) \sim \delta C_{\pi\pi}({\bf x}_{\rm ref},t) .
 \label{eq:condition_ratio_HL}
\end{equation}
This is a sufficient condition of flat $t$ dependence of $H_L(t,k;k) / C_{\pi\pi}({\bf x}_{\rm ref},t)$.

\begin{figure}[t]
 \centering
 \includegraphics[scale=0.60]{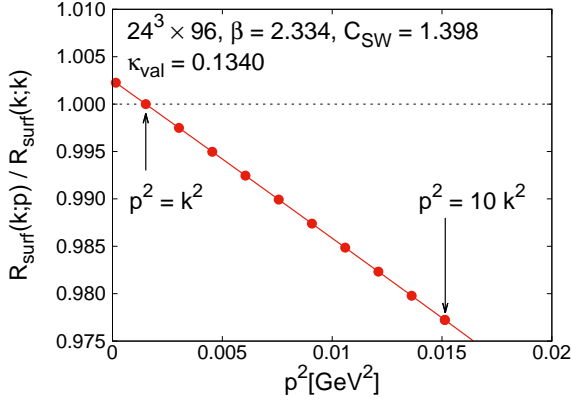}
 \caption{
  \label{fig:p2-ratio_surf_over_G}
  $p^2$ dependence of the ratio of
  $R_{\rm surf}(k;p) / R_{\rm surf}(k;k)$
  with ${\bf x}_{\rm ref} = (12,7,2)$.
 }
\end{figure}

We demonstrate the sufficient condition is realized,
if $C_{\pi\pi}({\bf x}_{\rm ref},t)$ is dominated by scattering states with almost zero momentum.
In the small $t$ region, $C_{\pi\pi}({\bf x},t)$ includes large contributions from
not only the ground state of two pions, but also scattering states
with the first radial excited state of $\pi$, denoting $\pi^\prime$.
We restrict our consideration below the energy of $\pi^\prime \pi^\prime \to \pi^\prime \pi^\prime$ scattering,
neglecting inelasticities.
Then, $C_{\pi\pi}({\bf x},t)$ is expressed as
\begin{eqnarray}
 C_{\pi\pi}({\bf x},t)
 &=& \sum_q A_q(t) \phi({\bf x};q)
 \nonumber \\
 &+& \sum_{q^\prime}
     A^\prime_{q^\prime}(t)
     \phi^\prime({\bf x};q^\prime)
 \nonumber \\
 &+& \sum_{q^{\prime\prime}}
     A^{\prime\prime}_{q^{\prime\prime}}(t)
     \phi^{\prime\prime}({\bf x};q^{\prime\prime}) ,
 \label{eq:C_pipi_exp}
\end{eqnarray}
where $A_q(t) = C_q e^{-E_q t}$,
$A^\prime_{q^\prime}(t) = C^\prime_{q^\prime}e^{-E^\prime_{q^\prime}t}$ 
and $A^{\prime\prime}_{q^{\prime\prime}}(t) = C^{\prime\prime}_{q^{\prime\prime}}e^{-E^{\prime\prime}_{q^{\prime\prime}}t}$
with $C_{q}$, $C^\prime_{q^\prime}$ and $C^{\prime\prime}_{q^{\prime\prime}}$ are overall constants for each contribution,
and
\begin{eqnarray}
 E_q
 &=& 2 \sqrt{m_\pi^2 + q^2} ,
 \\
 E^\prime_q
 &=& \sqrt{m_\pi^2 + q^2} + \sqrt{m_{\pi^\prime}^2+q^2} ,
 \\
 E^{\prime\prime}_q
 &=& 2\sqrt{m_{\pi^\prime}^2+q^2} .
\end{eqnarray}
The terms with the prime ($^\prime$) and double prime ($^{\prime\prime}$) correspond to 
contributions of scatterings for $\pi \pi^\prime \to \pi \pi^\prime$ and $\pi^\prime \pi^\prime \to \pi^\prime \pi^\prime$.
Substituting Eq.~(\ref{eq:C_pipi_exp}) to Eq.~(\ref{eq:HL_tkk}) provides
\begin{eqnarray}
 H_L(t,k;k)
 &=& \sum_q A_q(t) \mbox{surf}(k;q) 
 \nonumber \\
 &+& \sum_{q^\prime}
     A^\prime_{q^\prime}(t)
     \mbox{surf}^{\prime}(k;q^{\prime}) 
 \nonumber \\
 &+& \sum_{q^{\prime\prime}}
     A^{\prime\prime}_{q^{\prime\prime}}(t)
     \mbox{surf}^{\prime\prime}(k;q^{\prime\prime}) ,
 \label{eq:HL_tkk_with_surf}
\end{eqnarray}
where we use $\mbox{surf}(p;k)$ in Eq.~(\ref{eq:surface_j0}) in conjunction with $\Delta j_0(k x) = -k^2 j_0(k x)$,
ignoring the lattice artifact for simplicity,
\begin{eqnarray}
 - \sum_{\bf x} (\Delta + k^2) \phi({\bf x};q) j_0(k x)
 = \mbox{surf}(k;q).
\end{eqnarray}
The excited state contributions $\delta H_L(t,k;k)$ and $\delta C_{\pi\pi}({\bf x}_{\rm ref},t)$
for the numerator and denominator are expressed as
\begin{widetext}
\begin{eqnarray}
 \delta H_L(t,k;k)
 &=& \frac{
       \displaystyle{
           \sum_{q\ne k} A_q(t) \mbox{surf}(k;q)
         + \sum_{q^\prime} A^\prime_{q^\prime}(t) \mbox{surf}^{\prime}(k;q^{\prime})
         + \sum_{q^{\prime\prime}} A^{\prime\prime}_{q^{\prime\prime}}(t)
           \mbox{surf}^{\prime\prime}(k;q^{\prime\prime})
       }
     }
     {A_k(t) \mbox{surf}(k;k)},
 \label{eq:del_Htkk}
 \\
 \delta C_{\pi\pi}({\bf x}_{\rm ref},t)
 &=& \frac{
       \displaystyle{
           \sum_{q \ne k} A_q(t) \phi({\bf x}_{\rm ref};q) 
         + \sum_{q^\prime} A^\prime_{q^\prime}(t)
                           \phi^\prime({\bf x}_{\rm ref};q^\prime)
         + \sum_{q^{\prime\prime}} A^{\prime\prime}_{q^{\prime\prime}}(t)
                                   \phi^{\prime\prime}({\bf x}_{\rm ref};q^{\prime\prime})
       }
     }
     {A_k(t) \phi({\bf x}_{\rm ref};k)} .
 \label{eq:del_C_x_t}
\end{eqnarray}
\end{widetext}
The cancellation condition of Eq.~(\ref{eq:condition_ratio_HL}) implies
the coefficients of each state contribution in $\delta H_L(t,k;k)$ and $\delta C_{\pi\pi}({\bf x}_{\rm ref},t)$ coincide.
At some momentum $p$, the relation between the coefficients is
\begin{equation}
 \frac{\mbox{surf}(k;p)}{\mbox{surf}(k;k)}
 \sim
 \frac{\phi({\bf x}_{\rm ref};p)}{\phi({\bf x}_{\rm ref};k)} .
 \label{eq:condition_cancellation_of_H_kk_over_C_pi_pi}
\end{equation}
$\phi({\bf x}_{\rm ref};p)$ for $x_{\rm ref} > R$ can be expressed
by the solution of the Helmholtz equation $G({\bf x};p)$ defined in Eq.~(\ref{eq:expand_phi_G}).
The surface terms are also evaluated by $G({\bf x};p)$ through Eq.~(\ref{eq:surface_j0}),
supposing the surface boundary lies outside the interaction range $R$.
Substituting $G({\bf x};p)$ for $\phi({\bf x}_{\rm ref};p)$ and $\mbox{surf}(k;p)$ in $p^2 \sim k^2$ leads to
\begin{equation}
 \frac{\mbox{surf}(k;k)}{G({\bf x}_{\rm ref};k)}
 \sim
 \frac{\mbox{surf}(k;p)}{G({\bf x}_{\rm ref};p)} .
 \label{eq:surf_relation}
\end{equation}
Equation~(\ref{eq:surf_relation}) is numerically estimated.
We define a ratio $R_{\rm surf}(k;p)$ for the estimation,
\begin{equation}
 R_{\rm surf}(k;p)
 = \frac{\mbox{surf}(k;p)}{G({\bf x}_{\rm ref};p)}.
\end{equation}
We investigate $p^2$ dependence of $R_{\rm surf}(k;p)$ in the range of $0.1 k^2 \leq p^2 \leq 10 k^2$,
supposing unmeasured $q_0^\prime$ and $q_0^{\prime \prime}$ are in this range.
The result is plotted in Fig.~\ref{fig:p2-ratio_surf_over_G}.
Difference of $R_{\rm surf}(k;p)$ is less than 3 \% even at $p^2 = 10 k^2$.
Our data support validity of Eq.~(\ref{eq:condition_cancellation_of_H_kk_over_C_pi_pi}).

In summary, the excited states contamination in the ratio of $H_L(k;k) / (C_k \phi({\bf x}_{\rm ref};k))$ is understood
under the following conditions.
\begin{itemize}
 \item Energy is below $\pi^\prime \pi^\prime \to \pi^\prime \pi^\prime$ scattering
       with no inelasticities.
 \item Contribution from higher momentum states is small,
       due to our choice of the source operator.
 \item $q_0^\prime$ and $q_0^{\prime \prime}$ are assumed to be
       in the range of $0.1 k^2 \leq p^2 \leq 10 k^2$.
\end{itemize}
Then, contamination between $H_L(t,k;k)$ and $C_{\pi\pi}({\bf x}_{\rm ref},t)$ can be explained.
For further analysis, we need the variational method to distinguish the excited states.

\bibliographystyle{apsrev4-1}
\bibliography{ver_submit}

\end{document}